\documentclass[onecolumn,compsoc]{IEEEtran}
\usepackage{amsmath,amssymb,amsfonts}
\usepackage{algorithmic}
\usepackage{graphicx}
\usepackage{textcomp}
\usepackage{xcolor}
\usepackage{multirow}
\usepackage[nolist,nohyperlinks]{acronym}
\usepackage{url}
\usepackage{soulutf8}
\usepackage{enumerate}
\usepackage{svg}
\usepackage{url}
\usepackage{enumitem}
\usepackage{fontawesome}
\usepackage[hidelinks]{hyperref}
\usepackage[numbers]{natbib}
\usepackage[utf8]{inputenc}
\usepackage[T1]{fontenc}
\usepackage{zlmtt}
\usepackage{fancyvrb}

\usepackage{amssymb}

\usepackage{color}

\counterwithin{paragraph}{subsection}

\newcommand{\orcid}[1]{\href{https://orcid.org/#1}{\includegraphics[width=10pt]{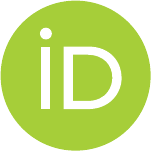}}}

\def\BibTeX{{\rm B\kern-.05em{\sc i\kern-.025em b}\kern-.08em
    T\kern-.1667em\lower.7ex\hbox{E}\kern-.125emX}}
    
\begin{document}

\begin{acronym}

  \acro{IoT}{\emph{Internet of Things}}
         \acro{PKI}{\emph{Public Key Infrastructure}}

     \acro{PoW}{\emph{Proof of Work}}

     \acro{GDPR}{\emph{General Data Protection Regulation}}

     	\acro{UTXO}{\emph{Unspent Transaction Output}}
     	
     	   \acro{RSA}{\emph{Rivest-Shamir-Adleman}}

     \acro{PoS}{\emph{Proof of Stakes}}
     
        \acro{PoA}{\emph{Proof of Authority}}

    \acro{IPFS}{\emph{InterPlanetary File System}}

\end{acronym}


\title{Analyzing the Impact of Elusive Faults on Blockchain Reliability
}

\author{\
\IEEEauthorblockN{
Fernando Richter Vidal\orcid{0000-0003-4869-2336}, Naghmeh Ivaki\orcid{0000-0001-8376-6711}, Nuno Laranjeiro\orcid{0000-0003-0011-9901}
}

\IEEEauthorblockA{
University of Coimbra\\
	Centre for Informatics and Systems of the University of Coimbra\\
	Department of Informatics Engineering, Portugal \\
		fernandovidal@dei.uc.pt,
		naghmeh@dei.uc.pt, cnl@dei.uc.pt
}
} 

\maketitle

\begin{abstract}

Blockchain recently became very popular due to its use in cryptocurrencies and potential application in various domains (e.g., retail, healthcare, insurance). The smart contract is a key part of blockchain systems and specifies an agreement between transaction participants. Nowadays, smart contracts are being deployed carrying residual faults, including severe vulnerabilities that lead to different types of failures at runtime. Fault detection tools can be used to detect faults that may then be removed from the code before deployment. However, in the case of smart contracts, the common opinion is that tools are immature and ineffective. In this work, we carry out a fault injection campaign to empirically analyze the runtime impact that realistic faults present in smart contracts may have on the reliability of blockchain systems. We place particular attention on the faults that elude popular smart contract verification tools and show if and in which ways the faults lead the blockchain system to fail at runtime. Results show general poor detection and, to some extent, complementary performance by the three tools used. The results also show that several elusive faults are responsible for severe blockchain failures.

\end{abstract}



\section{Introduction}
\label{sec:int}

Blockchain systems can be described as an implementation of a distributed ledger, in which transaction records are stored and linked together using a cryptography method. Such systems are supported by a peer to peer network, in which peers can join and obtain a copy of the state of the blockchain. When a new transaction arrives, a consensus protocol is used among the system peers that, by consensus, will accept or reject the transaction \cite{yaga_blockchain_2018}.

At the core of a blockchain system, we find the smart contract, a programmed application stored and executed by the blockchain middleware that contains the logic pertaining to a certain transaction, including pre-conditions that should be fulfilled so that the transaction concludes successfully  \cite{alharby_blockchain-based_2017}. Once a smart contract is deployed on the blockchain, it cannot be modified. Indeed, a faulty contract can only be terminated, and a new one must be put in place, which may have serious consequences (e.g., security, financial) for the involved parties \cite{praitheeshan_security_2020}.

Smart contracts are being deployed carrying residual bugs, including severe security vulnerabilities. This is due to several reasons, including a lack of developer expertise on the blockchain, the use of new programming languages (e.g., Solidity is a popular choice for programming smart contracts), the use of new coding tools, and the lack of mature defect verification tools. All of these factors combine and lead to the deployment of faulty contracts, which, at some point in time, see their faults being activated with diverse consequences, ranging from performance, ledger integrity violation, increased resource usage leading to gas depletion, among many others.

In this paper, we carry out a fault injection campaign to show the impact realistic faults can have on the reliability of a blockchain system, with particular attention to the types of faults that may elude popular smart contract verification tools. Notice that we use the terms \textit{blockchain reliability} to generally refer to issues that represent a deviation from the correct execution of transactions. In particular, we refer to the occurrence of the following types of issues and failures that : \textit{revert}, \textit{abort}, \textit{out-of-gas}, \textit{correctness}, \textit{integrity} and \textit{latent integrity}. By \textit{elusive} we refer to the particular cases of faults that escape detection by smart contract verification tools and whose impact we analyze by the end of this paper. In summary, our proposal includes going through the following three steps:

\begin{enumerate}
    \item \textit{Step 1:} We begin by resorting to an existing fault model that we have created in a previous study \cite{prdc2021}, based on faults observed in real smart contracts in the field (i.e., in this sense, representative of issues occurring in the context of blockchain), and implement injectors for a total of 36 faults.
    \item \textit{Step 2:} We inject faults in a set of 400 smart contracts, which have been randomly extracted from \cite{Ferreira2020}, to achieve a total of 15,494 faulty contracts (Injection of each fault may result in more than one faulty contract as some faults could be injected in several points in a contract.). We then deploy the original contracts and the faulty ones and run them against user transactions (a total of $10,925,749$ transactions are executed in this study over several months) so that we respectively understand expected/normal behavior and possible deviations (abnormal behavior), namely failures (e.g., aborted transactions, ledger integrity corruption) and performance degradation  (e.g., high memory consumption and high CPU usage).
    \item \textit{Step 3:} We map the observations to the fault detection capabilities of three state of the art fault detection tools, namely Mythril version v0.22.19 \cite{consensys_mythril_2021}, Slither v0.8.0 \cite{feist_slither_2019}, and Securify v0.0.1 \cite{tsankov_securify_2018}. The goal is to illustrate the overall effectiveness of the tools and especially signal the types of faults that tend to not be detected by the verification tools and their impact on the blockchain system.
\end{enumerate}

The main results of this work show that: i) the injected faults are capable of generating diverse types of failures at runtime; ii) the verification tools show low effectiveness, even in faults that seem to be easy to detect (e.g., \textit{Missing Compiler Version}) and, to some extent, their complementary nature; iii) there are indeed elusive faults, of which some are strongly connected to severe types of failures in blockchain systems, including \textit{latent integrity}, \textit{integrity}, and \textit{correctness} failures. The contributions of this paper are as follows:

\begin{itemize}
    \item The implementation of a set composed of 36 faults that are specific of smart contracts, made available via a fault injection tool at \cite{results}.
    \item A large-scale analysis of the impact of different types of realistic faults injected on an initial set of 400 real smart contracts and resulting in 15,494 mutated contracts (i.e., For every single fault, each contract has at least one faulty mutant).
    \item The evaluation of the fault detection effectiveness of three state of the art smart contract verification tools (i.e., Securify2 \cite{tsankov_securify_2018}, Mythril \cite{consensys_mythril_2021}, and Slither \cite{feist_slither_2019}), in presence of the different types of faults.
    \item An analysis of the impact of faults that tend to elude verification tools on the reliability of the blockchain system.
    
\end{itemize}

This paper is organized as follows. Section II presents background and related work and Section III presents the design of our experimental study. Section IV discusses the results and Section V presents the main threats to the validity of this work. Finally, Section VI concludes this paper.


\section{Background and Related Work}
\label{sec:relatedwork}

This section first presents background on smart contracts security issues. It then reviews the security and verification tools and techniques. Finally, it presents the work evaluating the security and verification tools.  

\subsection{Smart Contracts' Security Vulnerabilities}

To characterize the vulnerabilities in smart contracts, there are mostly two classifications, namely Decentralized Application Security Project (DASP) \cite{DASP} and Smart Contract Weakness Classification (SWC) \cite{SWC}. DASP is a collaborative project of the NCC Group \cite{DASP} to characterize the smart contract vulnerabilities. It includes 9 classes of vulnerabilities. SWC is a classification supported by the Mythx team to characterize the smart contract vulnerabilities and is based on Common Weakness Enumeration (CWE). It includes 37 classes of vulnerabilities.

In addition to these two classifications, we can find specific sets of vulnerabilities (including other types) in some research papers. \citeauthor{harz_towards_2018} \cite{harz_towards_2018} present a survey focused on the security aspect of smart contract programming languages. It provides an overview of the current programming languages for implementing smart contracts (a total of 19 programming languages), their security features, and other information like the paradigm, instruction set, semantics, and metering.  \citeauthor{praitheeshan_security_2020} \cite{praitheeshan_security_2020} present a list of 16 Ethereum smart contracts vulnerabilities and 19 software security issues.  Peng et al. \cite{Peng_2021} review the use of smart contracts in the context of IoT applications, mentioning the main security challenges due to the vulnerabilities identified in the programs. The findings reported by the authors is that real-world applications are still in their infancy, and despite the security auditing tools that exist, they can only detect a fraction of known vulnerabilities. 

\subsection{Smart Contract Security Assessment Tools and Techniques}

In this section, we analyze the techniques and tools identified during our review of state of the art in smart contract security verification and security assessment. The identified works fall into four main categories: \textbf{formal verification techniques} \cite{Bhargavan2016}, \textbf{static analysis techniques} \cite{Chess2004}, \textbf{software testing techniques}, and \textbf{machine learning-based techniques} \cite{Momeni2019}. 

The Formal Verification category includes methods based on formal proofs or abstract mathematical models of a certain system or part of the system to prove its correctness (e.g., mainly functional correctness) \cite{Seligman2015}. The specific techniques in this category are Model Checking \cite{kalra_zeus_2018,M.Clarke1999} and Theorem Proving \cite{Schumann2001}.

The Static Analysis category includes methods that do not require the code to be executed and rely on the inspection of code by various means (e.g., pattern recognition, taint analysis) to discover software defects \cite{Rival2016}.  Specific techniques in this category are Abstract Interpretation \cite{AbstractInterpretation2021}, Taint Analysis \cite{Xu2018}, Pattern Recognition \cite{zhang_soliditycheck_2019}.

The Software Testing category includes methods that rely on software execution with the intention of finding defects \cite{Myers2012}. The specific techniques in this category are Fuzzing \cite{Chen2018}, Mutation Testing \cite{Wong2001}, or Symbolic Execution \cite{Baldoni2019, luu_making_2016, Sunbeom2021}.

The Machine Learning category includes artificial intelligence-based techniques that focus on the study of algorithms that can learn from experience and historical data to detect anomalies or predict new output values (i.e., vulnerabilities). In this context, most techniques are based on supervised learning, which involves using labeled datasets to train algorithms that classify data or predict outcomes for a particular output. The specific techniques in this category are Classical Machine Learning Models: classical machine learning models \cite{wang_contractward_2020}, Deep Learning Models \cite{Zhuang2020GraphNeural}, and Ensemble Learning Models \cite{Novel2022}. Notice that some tools use more than one technique to achieve their goals. For instance, Smartian \cite{Choi2021a} uses a combination of static and dynamic analysis techniques for fuzzing smart contracts.



At this point, we highlight three verification tools, namely Slither, Securify, and Mythril, which are subject of analysis in this work. \textbf{Slither} \cite{feist_slither_2019} is a static analysis tool that uses taint analysis. The tool compiles the Solidity smart contract source code into a Solidity Abstract Syntact Tree (AST) to extract the contract's inheritance graph, the control flow graph (CFG), and the list of expressions. Then it transforms the code of the contract to SlithIR, its internal representation language, that uses static single assessment (SSA) to facilitate the computation of a variety of code analyses. It also includes a graph for code understanding and assisted code review. The authors evaluated the tool in 1,000 most used smart contracts (that had the largest number of transactions) to find that it outperforms three other popular tools (Solhint \cite{Solhint}, SmartCheck \cite{tikhomirov_smartcheck_2018}, Securify \cite{tsankov_securify_2018}) in terms of performance, robustness (i.e., whether the tool completed the execution), and accuracy (i.e., false positives reported).

\textbf{Securify} \cite{tsankov_securify_2018} is a static analysis tool that uses Soufflé, a scalable Datalog solver, to symbolically analyze the EVM bytecode of the smart contract and extract semantic facts, and then checks those semantics against violation patterns. Thus, the tool was developed based on a set of compliance and violation security patterns that capture sufficient conditions to prove and disprove practical security properties. To foster extensibility, the patterns are specified in a language that is domain-specific.

\textbf{Mythril} is an open-source tool for analyzing the security of smart contracts, based on symbolic execution, SMT solving and taint analysis. It is able to detect software defects, namely various security vulnerabilities in smart contracts implemented for Ethereum and several other EVM-compatible blockchains. The tool is often found being used in research related with smart contract evaluation, e.g., \cite{ghaleb_how_2020}, \cite{parizi_empirical_2018}, \cite{akca_solanalyser_2019}.

\subsection{Verification Tools Assessment}

In \cite{parizi_empirical_2018} the authors evaluate Oyente, Securify, Mythril and Smartcheck using ten contracts and express the tools' effectiveness by performing a Receiver Operating Characteristic (ROC) analysis and also analyze accuracy, revealing differences and gaps in the tools' effectiveness. A framework for analyzing and testing smart contracts is presented in \cite{akca_solanalyser_2019}. The authors evaluate their proposal with Oyente, Securify, Maian, SmartCheck and Mythril against 1,838 contracts and 8 faults which are used to produce 12,866 mutated contracts. Precision and recall are used to characterized the tools detection capabilities. 

A bug benchmark is proposed in \cite{ye_towards_2019}, which is then demonstrated by running Oyente and Slither against 1,010 contracts randomly selected from \texttt{etherscan.io}, highlighting the detection deficiencies of the tools when in presence of well-known vulnerabilities. The authors in \cite{ghaleb_how_2020} evaluate the bug detection effectiveness of several static analysis tools, namely Oyente, Securify, Mythril, Smartcheck, Manticore, and Slither. The approach is based on the injection of bugs in contracts, based on known bug patterns. Injection is performed with types of bugs that the tools indicate are able to detect and use typical metrics, such as false-negative and false-positive rates to assess the tools' performance.

Nine analysis tools for smart contracts are evaluated in \cite{durieux_empirical_2020}. The authors use 47,587 Ethereum smart contracts, highlighting clear deficiencies in the tools detection capabilities, including the tool marked as most accurate tool (Mythril), able to detect only 27\% of the vulnerabilities. In \cite{Ren2021}, the authors present an empirical evaluation of 9 contract verification against 46,186 smart contracts. Main findings include the recommendation of a set of diverse test suites; a unified execution environment with suitable runtime parameters; and more quantitative and multi-dimensional performance metrics.


We, in a previous work \cite{prdc2021}, presented a fault-injection approach to analyze the effectiveness of three static verification tools, Mythril, Securify, and Slither. However, the work was a proof of concept, and the study was limited to a small number of contracts and a small number of vulnerabilities. In this work, we consider a large number of smart contracts with diverse types of vulnerabilities. 

Moreover, the works mentioned above for evaluation of verification tools generally share the view that smart contract verification tools are immature, reflected in their detection capabilities. However, such a vision is generally not put in perspective with the runtime effect of smart-contract specific faults. We, in another previous work \cite{hajdu_using_2020}, tried to analyze the effect of faults on blockchain systems, but it was done on a very small scale (with only 5 contracts). The other studies tend not to analyze in depth the faults (and their effects) that elude smart contract verification tools, which is one of the main objectives of this work.


\section{Approach}
\label{sec:approach}

This section presents the approach followed in this work. The next subsections go through the following main steps:

\begin{enumerate}
    \item The fault injection approach, which consists of the injection of faults in a set of smart contracts;
    \item The procedure for analyzing the impact the injected faults have on each contract;
    \item The procedure to analyze the effectiveness of smart contract fault detection tools;
    \item The analysis of the impact of faults that tend to elude smart contract verification tools.
\end{enumerate}

\subsection{Fault Injection Approach}

The starting point for this work is the ability to inject faults in smart contracts. For this purpose, we follow the long-established tradition of software fault injection in which, based on a model that represents real faults (i.e., faults observed in real systems in the field), 'probable' software faults are artificially introduced in a certain component of a larger system \cite{Natella2016}. This allows to understand the effect that a certain type of fault can have on a system, once it is activated (e.g., does the system fail catastrophically, does it have its performance degraded) \cite{faultinjection}. This type of technique can be towards various goals, namely for evaluating systems behavior \cite{Natella2016}, test suite effectiveness (i.e., in the case of mutation testing approaches) \cite{marques_injecting_2021}, tools that act over the systems, e.g., vulnerability scanners \cite{webscanners} or even for failure prediction \cite{failureprediction}. We apply a software-implemented fault injection (SWIFI) technique \cite{Natella2016}, which we have succesfully used in the past \cite{hajdu_using_2020}, although in a much narrower scope (a different set of faults was used in 5 contracts). For completeness, we conceptually overview the technique in the next paragraphs.

\begin{figure}[ht]
\centering
\includegraphics[scale=0.59]{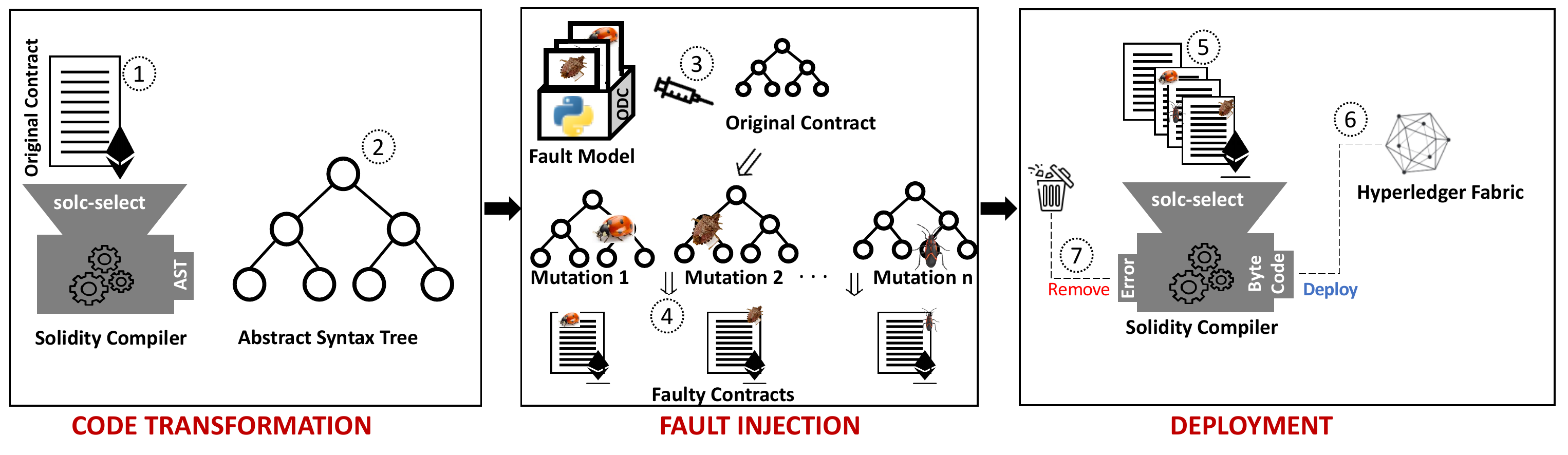}
\caption{Fault injection process.}
\label{fig:fault-injection-process}
\end{figure}

Figure \ref{fig:fault-injection-process}  presents our fault injection process. The first phase (on the left side of the image) consists of transforming the original code (i.e., solidity format) into AST format. To perform this task, we use native solidity compiler functions (i.e., \textit{solc} with \textit{–ast-json} argument) to first compile the code (step 1) and then transform it into AST (step 2). This way, AST can be generated only for the contracts that are successfully compiled. 

The second phase (in the middle) transforms the previously created ASTs into faulty smart contracts (step 3). For each fault existing in our fault model, at least one faulty contract will be generated for a given AST. This is because a fault can be injected in different forms (e.g., a wrong arithmetic expression may take various forms) or injected in different places within the AST. Step 3 results in a set of faulty ASTs. In the next step, step 4, all faulty ASTs are transformed back into their original format (i.e., solidity code). To achieve this we implemented the necessary code for applying the transformation and verified its correctness by manually inspecting the resulting file and comparing it to the original code. Two Early State Researchers were involved in this verification process. This final step ends up in a set of faulty contracts called mutants. 

The last phase is focused on the deployment of the mutants. In step 5, we verify whether the mutants are still valid executable programs (i.e., by compiling them). The successfully compiled programs are then deployed into a Hyperledger fabric (step 6), and the remaining contracts are removed from the analysis (step 7). 

Figure \ref{fig:code-example} shows an example with a contract (on the left-hand side, named original) and the corresponding faulty version (on the right-hand side). 
The faulty version represents a mutation of the original contract that holds a specific vulnerability named \textit{Uninitialized Storage Pointer} (SWC-109) \cite{SWC}. This vulnerability refers to a situation where a local storage variable remains uninitialized and may be used to point to an unexpected storage location in the contract. This may lead to an unexpected behavior either intentionally caused by an attacker or unintentionally. The program allows authorized people to transfer money to suppliers and has been inspired by the examples in \cite{antonopoulos_mastering_2018}. The concurrent payments are controlled by the \textit{unlocked} variable and \textit{require} instruction that only allows payments when the contract is unlocked.  In the original contract, there is a local variable, namely \textit{newTransfer}, that is appropriately initialized 
by marking the variable's storing location explicitly with the \textit{memory} attribute (line 8). In contrast, the faulty contract does not explicitly mark its storing location, and the variable remains uninitialized as a local variable (it is instead a global storage variable).


\begin{figure*}
\centering
\vspace{0.5cm}
\begin{minipage}{0.42\linewidth}
\begin{Verbatim}[frame=topline,numbers=left,label=Original Contract,framesep=3mm,commandchars=\\\{\}]
contract PaySupplier \{
   bool public unlocked = false;
   
   ...
   
   function TransferMoney(bytes32 _name) 
   public \{
      Person \textcolor{blue}{memory} newTransfer;
      newTransfer.name = _name;
      
      ...
      
      require(unlocked); 
    \}
\}
\end{Verbatim}
\end{minipage}\hfill
\begin{minipage}{0.42\linewidth}
\begin{Verbatim}[frame=topline,numbers=left,label=Faulty Contract,framesep=3mm,commandchars=\\\{\}]
contract PaySupplier \{
   bool public unlocked = false;
   
   ...
   
   function TransferMoney(bytes32 _name) 
   public \{
      Person \textcolor{red}{storage} newTransfer;
      newTransfer.name = _name;
      
      [several lines of code]
      
      require(unlocked); 
    \}
\}
\end{Verbatim}
\end{minipage}
\caption{A faulty contract example.}
\label{fig:code-example}
\end{figure*}

As discussed, in order to inject the SWC-109 vulnerability, we convert the original code into an AST representation and then inject the vulnerability. Figure \ref{fig:injector109} shows an illustrative Python example of the injector of this vulnerability. The injector searches the AST for a condition (implemented in function \texttt{condition}). When the condition matches, the node/attribute of the tree is localized, then the changes are applied (see function \texttt{changeTre}, resulting in a faulty AST. In this specific example, the injector looks for the \textit{memory} attribute used for initialization of a local storage variable and replaces it with "storage". At the end, the faulty AST is transformed back into the faulty contract code. The implementation of the whole set of 36 faults is available at \cite{results}.


\begin{figure}
\vspace{0.5cm}
\begin{minipage}{1.0\linewidth}
\begin{Verbatim}[frame=topline,numbers=left,label=Injector for \textbf{Uninitialized Storage Pointer} vulnerability,framesep=3mm]
from common import mainfunc

def condition(root, node):
    return node['attributes']["initialValue"] == None and node['name'] == 'VariableDeclaration'     
    and node['attributes']['storageLocation'] == "memory"

def changeTree(root, node):
    node['attributes']['storageLocation'] = "storage"
    node['attributes']['type'] = node['attributes']['type'].replace("memory", "storage")

if __name__ == "__main__":
    mainfunc('Removes memory initialization pointers.', condition, changeTree, '4')
\end{Verbatim}
\end{minipage}\hfill 

\caption{Code example for the injection of the SWC-109 vulnerability.}
\label{fig:injector109}
\end{figure}

After injecting the faults, we will have a list of fault-free smart contracts (i.e., the original contracts, without known faults) and their corresponding faulty smart contracts to be used in our evaluation, which is composed of the following studies:

\begin{itemize}
    \item \textbf{Study 1 - Faults' Impact}: we evaluate the behavior of the blockchain system in presence of the injected faults. For this, we execute both fault-free and their respective faulty versions individually on an isolated environment of the blockchain. We compare the outcome of the fault-free runs and the faulty runs. By having the fault-free runs we have a reference to evaluate the impact of each fault.
    
    \item \textbf{Study 2 - Effectiveness of Verification Tools}: we evaluate the effectiveness of smart contract verification tools, namely the tools fault detection capabilities against a set of faulty contracts generated by the fault injection tool, based on the faults described on our fault model.
    
    \item \textbf{Study 3 - Impact of Elusive Faults}: We analyze the impact of the faults that tend to escape detection by the verification tools.
    
\end{itemize}

\subsubsection{Fault Model}

In this work, we opted to resort to an existing fault model created in our previous work \cite{prdc2021}, 
for selecting smart contract-specific faults (for implementation in the fault injection process). The defects are organized based on the Orthogonal Defect Classification (ODC) defect classification scheme \cite{ibm_orthogonal_2013} and we tried to implement at least one fault for each ODC class and each defect type, balancing, at the same time, the effort required (some faults of the same class and type are very similar, and, in this sense, trivial to implement). We reached a total of 36 implemented faults, which we present in Table \ref{tab:scheme}.



\renewcommand{\arraystretch}{1.2}
\begin{table}[ht]
\caption{Smart contract defect classification \cite{prdc2021}}
\label{tab:scheme}
\resizebox{\textwidth}{!}{
\begin{tabular}{|l|l|l|l|}
\hline
\textbf{Defect   Class}      & \textbf{Defect Nature}   & \textbf{Defect Name}                                                                  & \textbf{Defect   Identifier} \\
\hline \hline
{Assignment} & {Missing} & Initialization of Storage Variables/Pointers (Uninitialized Storage Pointer)   \textbf{(MISP)} & A\_MISP                       \\ \cline{3-4} 
                             &                          & Initialization of Local Variable \textbf{(MILV)}                                               & A\_MILV                       \\ \cline{3-4} 
                             &                          & Initialization of State Variables \textbf{(MISV)}                                              & A\_MISV                       \\ \cline{3-4} 
                             &                          & Constructor \textbf{(MC)}                                                                      & A\_MC                         \\ \cline{3-4} 
                             &                          & Compiler Version \textbf{(MCV)}                                                                & A\_MCV                        \\ \cline{2-4} 
                             & {Wrong}   & Arithmetic Expression Used In Assignment \textbf{(WVAE)}                                       & A\_WVAE                       \\ \cline{3-4} 
                             &                          & Integer Sign \textbf{(WIS)}                                                                    & A\_WIS                        \\ \cline{3-4} 
                             &                          & Integer Truncation \textbf{(WIT)}                                                              & A\_WIT                        \\ \cline{3-4} 
                             &                          & Value Assignment With Too Many Digits \textbf{(WVATMD)}                                        & A\_WVATMD                     \\ \cline{3-4} 
                             &                          & Value Assigned To Contract Address \textbf{(WVAA)}                                             & A\_WVAA                       \\ \cline{3-4} 
                             &                          & Constructor Name \textbf{(WCN)}                                                                & A\_WCN                        \\ \cline{3-4} 
                             &                          & Variable Type (e.g., byte[]) \textbf{(WVT)}                                                    & A\_WVT                        \\ \cline{3-4} 
                             &                          & Declaration Of Invariant State Variable \textbf{(WDISV)}                                       & A\_WDISV                      \\ \cline{3-4} 
                             &                          & Variable Name (Variable Shadowing) \textbf{(WVN)}                                              & A\_WVN                        \\ \hline
{Checking}   & {Missing} & "require" On Transaction Sender \textbf{(MRTS)}                                                & CH\_MRTS                      \\ \cline{3-4} 
                             &                          & "require" On Input Variable(s) \textbf{(MRIV)}                                                 & CH\_MRIV                      \\ \cline{3-4} 
                             &                          & "require" OR Subexpression On Transaction Sender \textbf{(MROTS)}                              & CH\_MROTS                     \\ \cline{3-4} 
                             &                          & "require" OR Subexpression On Input Variable(s) \textbf{(MROIV)}                               & CH\_MROIV                     \\ \cline{3-4} 
                             &                          & "require" AND Subexpression On Transaction Sender \textbf{(MRATS)}                             & CH\_MRATS                     \\ \cline{3-4} 
                             &                          & "require" AND Subexpression On Input Variable(s) \textbf{(MRAIV)}                              & CH\_MRAIV                     \\ \cline{3-4} 
                             &                          & Check On Gas Limit \textbf{(MCHGL)}                                                            & CH\_MCHGL                     \\ \cline{3-4} 
                             &                          & Check On Arithmetic Operation \textbf{(MCHAO)}                                                 & CH\_MCHAO                     \\ \cline{3-4} 
                             &                          & Check On Suicide Functionality \textbf{(MCHSF)}                                                & CH\_MCHSF                     \\ \cline{2-4} 
                             & Wrong                    & "require" For Authorization (Authorization Through tx.origin) \textbf{(WRA)}                   & CH\_WRA                       \\ \hline
{Interface}   & {Missing} & Visibility modifier of state variables (implicit visibility) \textbf{(MVMSV)}                  & I\_MVMSV                      \\ \cline{3-4} 
                             &                          & Function Visibility Modifier \textbf{(MFVM)}                                                   & I\_MFVM                       \\ \cline{2-4} 
                             & Wrong                    & Visibility (public) for private/internal function \textbf{(WVPF)}                              & I\_WVPF                       \\ \hline
{Algorithm}   & {Missing} & "if" construct on transaction sender plus statements \textbf{(MITSS)}                          & AL\_MITSS                     \\ \cline{3-4} 
                             &                          & "if" construct on input variable(s) plus statements \textbf{(MIIVS)}                           & AL\_MIIVS                     \\ \cline{2-4} 
                             & {Wrong}   & Use of require, assert, and revert \textbf{(WRAR)}                                             & AL\_WRAR                      \\ \cline{3-4} 
                             &                          & Exception Handling \textbf{(WEH)}                                                              & AL\_WEH                       \\ \cline{2-4} 
                             & Extraneous               & Continue-statements in do-while-statements or for \textbf{(ECSWS)}                             & AL\_ECSWS                     \\ \hline
 {Function}    & {Missing} & Withdraw function \textbf{(MWF)}                                                               & F\_MWF                        \\ \cline{3-4} 
                             &                          & Inheritance \textbf{(MINHERITANCE)}                                                            & F\_MINHERITANCE               \\ \cline{2-4} 
                             & Wrong                    & Inheritance and inheritance Order \textbf{(WIO)}                                               & F\_WIO                        \\ \cline{2-4} 
                             & Extraneous               & Inheritance \textbf{(EINHERITANCE)}                                                            & F\_EINHERITANCE               \\ \hline
\end{tabular}
}
\end{table}

The set of 36 implemented faults is a subset of all faults that may affect a blockchain system (e.g., for the time being we did not implement reentrancy faults). Despite this, it is important to mention that the focus of this work is not on the definition of a fault model but is instead on the overall method proposed that ends up in the analysis of the effect of a subset of faults that tend to escape detection tools. In this sense, the specific faults used may vary as well as the specific tools used for this purpose.


\subsubsection{Smart Contracts Dataset}

In order to \textbf{identify a set of contracts} that could be used as input for these experiments, we randomly selected 400 of the contracts used in the work by Durieux et al.  \cite{durieux_empirical_2020} and that are available at 
\cite{results}. 
Next, each contract is passed to the fault injection tool, which determines which of the 36 faults can be injected and in how many code locations of that contract. Then the tool iteratively generates the respective fault contracts. This process resulted in a total of 15,494 faulty contracts (each faulty contract carries exactly one fault).

\subsection{Study 1 - Faults' Impact}
\begin{figure}[t]
    \centering
    \includegraphics[scale=0.6]{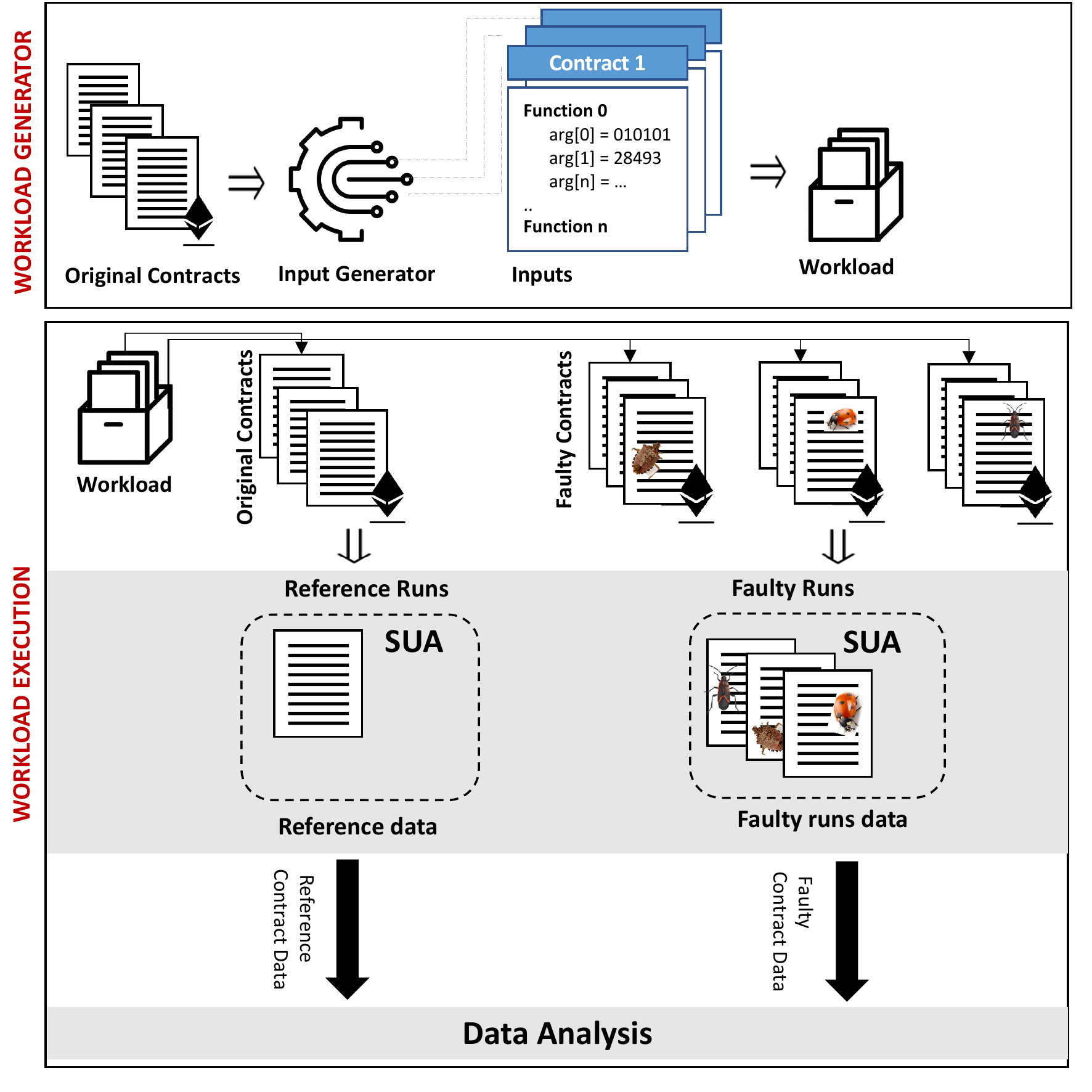}
    \caption{Approach for analyzing the faults' impact on the blockchain system.}
    \label{fig:experiment-1-approach}
\end{figure}

This section presents the approach followed to study the effect that our injected faults may have in smart contracts. Figure \ref{fig:experiment-1-approach} illustrates the approach, which in practice, consists of the following steps:

\begin{enumerate}
    \item Generation of the smart contracts' workload.
    \item Execution of both fault-free and faulty smart contracts in a private network.
    \item Analysis of the results, based on a set of metrics of interest.
\end{enumerate}

\textit{Step 1)} refers to the \textbf{smart contract workload generation}, which will allow activating the injected faults during the calls to the smart contract operations (i.e., the execution of transactions). We follow a simple workload generation procedure, with the goal of being able to execute most of the contracts and knowing that, in some cases, the generation of calls would require a more complex implementation of this procedure. We emphasize that the goal is to generate valid calls (as opposed by generating invalid or malicious calls, as for instance, a fuzzer would do). In practice, we  search for functions  in the smart contracts and generate values for the corresponding input parameters based on their \textit{type}, \textit{literals} that appear in the code, and \textit{randomly}, as follows:
    \begin{enumerate}
        \item \textit{Type-based:} We generate values according to the data type of the argument, e.g., true and false for booleans, minimum, maximum or zero for integers. Regarding arrays and strings, the values are recursively generated considering various lengths, including zero.
        
        \item \textit{Literal-based:} Literals present in the function are used as input for any arguments that match the literal type. This is done, as inputs are many times compared with literals, which then determines the flow inside the function code.
        
        \item \textit{Randomly}: Random values are also generated, namely for arguments of type Integer (within the minimum and maximum range) and Strings (with random characters and at random sizes). Arrays are generated with random elements of their type and also using different lengths.
    \end{enumerate}

We currently allow for up to 1500 function calls per function, which we found to be generally sufficient in terms of transaction diversity while maintaining the total number of transactions generated at reasonable levels (for data analysis). This workload generation process only takes place for the set of original smart contracts, and then each generated faulty contract is executed against the workload generated for the original contract. This way, we are able to compare the runtime behavior during the execution of the faulty contracts in contrast with its corresponding fault-free run (i.e., a golden run). The metrics considered for the comparison are discussed later in this subsection.

In \textit{Step 2}, we \textbf{execute the smart contracts in a private network}. In terms of environment we resort to a deployment of Hyperledger Fabric, with the Ethereum Virtual Machine (EVM) version of Hyperledger Burrow. We then use Hyperledger Caliper to perform the test runs, which is a blockchain benchmarking tool that allows users to measure the performance of a blockchain implementation against some predefined use cases \cite{noauthor_hyperledger_nodate}. The test runs are carried out by executing the respective workload in both the fault-free contracts and the corresponding generated faulty versions. The execution of a test run follows the next order:

\begin{enumerate}
    \item Hyperledger is set in a clean initial state, which means that the respective nodes (orderers and endorsers) are set (or reset) to an empty blockchain.
    \item The contract under evaluation is deployed onto the blockchain (in this case, onto the endorsing peers).
    \item The workload generated for the contract is executed and metrics about each transaction are collected.
\end{enumerate}

During the test runs, Caliper provides multiple transaction details, such as the timing of each transaction phase, side effects returned by the platform and other status information. In the end, the collected data is post-processed to match and compare the information of each transaction occurred in the faulty contracts with the corresponding transaction in the reference contracts.

The choice of the Hyperledger platform for the experimental setup is mostly related with the fact that it offers easy means to collect various metrics needed (e.g., transaction execution time, reverted transactions, CPU/memory usage). In what concerns performance metrics, notice that the goal is not to obtain absolute performance values, but to understand what is the relative impact in realistic conditions. Thus, the setup is similar to other studies where performance has been studied \cite{Choi2021, Mazzoni2022}. As we had the goal of creating an injector that is independent of the programming language, and due to the fact that Hyperledger includes modular blockchain frameworks, this decision of using Hyperledger is beneficial for future work, where the infrastructure may easily be reused to run programs in different blockchains.

After finishing the runs, in \textit{Step 3} we \textbf{analyze the results}. We compare the outcome of both the reference data (i.e., the outcome of the fault-free runs) and mutation data (i.e., the outcome of faulty smart contract runs).
For this, we consider the successful commit of the transactions performed in the test cases, as well as the differences and failures that arise in the transactions. In each execution, a \textit{transaction is only deemed successful} if i) all of its endorsements are successful and matching and ii) it is successfully ordered and reported as committed by all endorsing peers. In previous work we identified several different types of blockchain failures \cite{hajdu_using_2020}, which also fit the types of failures discussed in related work, e.g., \cite{Chacko2021}\cite{Chen2023}. Based on this, and in our own empirical analysis of the different failures during the experiments, we match our observations to following failure modes:

\begin{itemize}
    
    \item \textbf{Revert failure}: When Revert occurs, the execution of the transaction is stopped, and all state changes are rolled back. The reverted transaction consumes the gas used up to the point where the transaction is reverted. This failure mode, in our context, indicates whether there was at least one transaction in the faulty contract that was reverted while its reference instance did not. 

    \item \textbf{Abort failure}:Like \textit{Revert Failure}, when an abort occurs, the execution of the transaction is stopped, and all state changes are rolled back. The difference is that the aborted transaction consumes all gas up to the maximum allowance of the transaction. This failure mode indicates whether at least one transaction in the faulty contract was aborted while its reference instance did not fail.

    \item \textbf{Out-of-Gas failure}: Indicates whether there was at least one transaction in the faulty contract that failed due to gas depletion, while in its reference case it did not happen.
    
    \item \textbf{Correctness failure}: Indicates whether there was at least one transaction in the faulty contract that outputted a different result or return value than its reference fault-free contract. Correctness allows us to observe failures that can be seen by the client during output invariant checks.
    
    \item \textbf{Integrity failure}: Indicates whether there was at least one transaction in the faulty contract that outputted a different result or return value, and also a different read-write set than the reference fault-free contract (i.e., that transaction modified the state of the blockchain differently than the one from the reference contract). Integrity allows us to observe failures to the ledger integrity that can be seen by the client.
    
    \item \textbf{Latent integrity failure}: Indicates whether there was at least one transaction in the faulty contract that outputted the same result or return value than its reference fault-free contract, but with a different read-write set than the reference contract. The aim here is to observe errors that stay hidden and cannot be directly seen by the client, as it receives the expected result or return value.
\end{itemize}

Table \ref{table:analysis-failures} overviews the failure model considered in this work for analysis of the results. 

\begin{table}[h]
\centering
\caption{Failure modes and its characteristics}
\label{table:analysis-failures}
\begin{tabular}{|l|l|l|l|}
\hline
\textbf{Failure Modes}     & \textbf{\begin{tabular}[c]{@{}l@{}}Transaction \\ not concluded\end{tabular}} & \textbf{\begin{tabular}[c]{@{}l@{}}Incorrect \\ return value or \\ transaction result\end{tabular}} & \textbf{\begin{tabular}[c]{@{}l@{}}Incorrect \\ ledger state\end{tabular}} \\ \hline \hline
Abort            & \multicolumn{1}{c|}{\tiny\faCircle}                                                        &                                                                                                       &                                                                            \\ \hline
Revert           & \multicolumn{1}{c|}{\tiny\faCircle}                                                        &                                                                                                       &                                                                            \\ \hline
Out-of-gas       & \multicolumn{1}{c|}{\tiny\faCircle}                                                        &                                                                                                       &                                                                            \\ \hline
Correctness      &                                                                               & \multicolumn{1}{c|}{\tiny\faCircle}                                                                                &                                                                            \\ \hline
Integrity        &                                                                               & \multicolumn{1}{c|}{\tiny\faCircle}                                                                                & \multicolumn{1}{c|}{\tiny\faCircle}                                                     \\ \hline
Latent integrity &                                                                               &                                                                                                       & \multicolumn{1}{c|}{\tiny\faCircle}                                                     \\ \hline
\end{tabular}
\end{table}

In order to characterize the failures, we see whether the transaction is concluded, whether the result of a transaction (return value) is correct, and finally, whether the ledger state is correct. 
As shown in the table, when \textit{Abort and Revert failures} occur, neither a value (transaction result) is returned to the client nor any changes are made to the state of the ledger. The transaction fails, and some error or exception is delivered to the client. The only difference between the \textit {Abort Failure} and \textit{Revert Failure} is related to the gas consumption. A reverted transaction consumes the gas used up to the point where the transaction is reverted, while an aborted transaction consumes all gas up to the maximum allowance of the transaction.

In the case of \textit{Out-of-Gas failures}, similar to the previous cases, no value is returned, and no changes to the ledger state are made. However, the transaction is not concluded due to gas depletion (e.g., a fault may cause spending more resources).
In a \textit{Correctness failure}, the transaction is successfully concluded, but the transaction result is different from the reference run. In this case, the state of the ledger remains intact.
In contrast, in the case of \textit{Integrity failure}, in addition to having incorrect returned values, the integrity of the ledger's state is disrupted too. Finally, \textit{Latent integrity failure} relates to changes in the integrity of the ledger state, although the transaction result (values returned to the client) is correct (which means that a client cannot detect the problem). Although \textit{Correctness} and \textit{Integrity} failures are both severe, being undetectable makes the \textit{Latent Integrity Failure} the most severe failure in our failure model.  

\subsection{Study 2 - Effectiveness of Verification Tools}

This section presents the study for assessing the detection capabilities of the verification tools (i.e., Mythril, Securify2 and Slither), which is depicted in Figure \ref{fig:experiment-2-approach}. In practice, we go through the following steps:

\begin{figure}[th]
    \centering
    \includegraphics[scale=0.7]{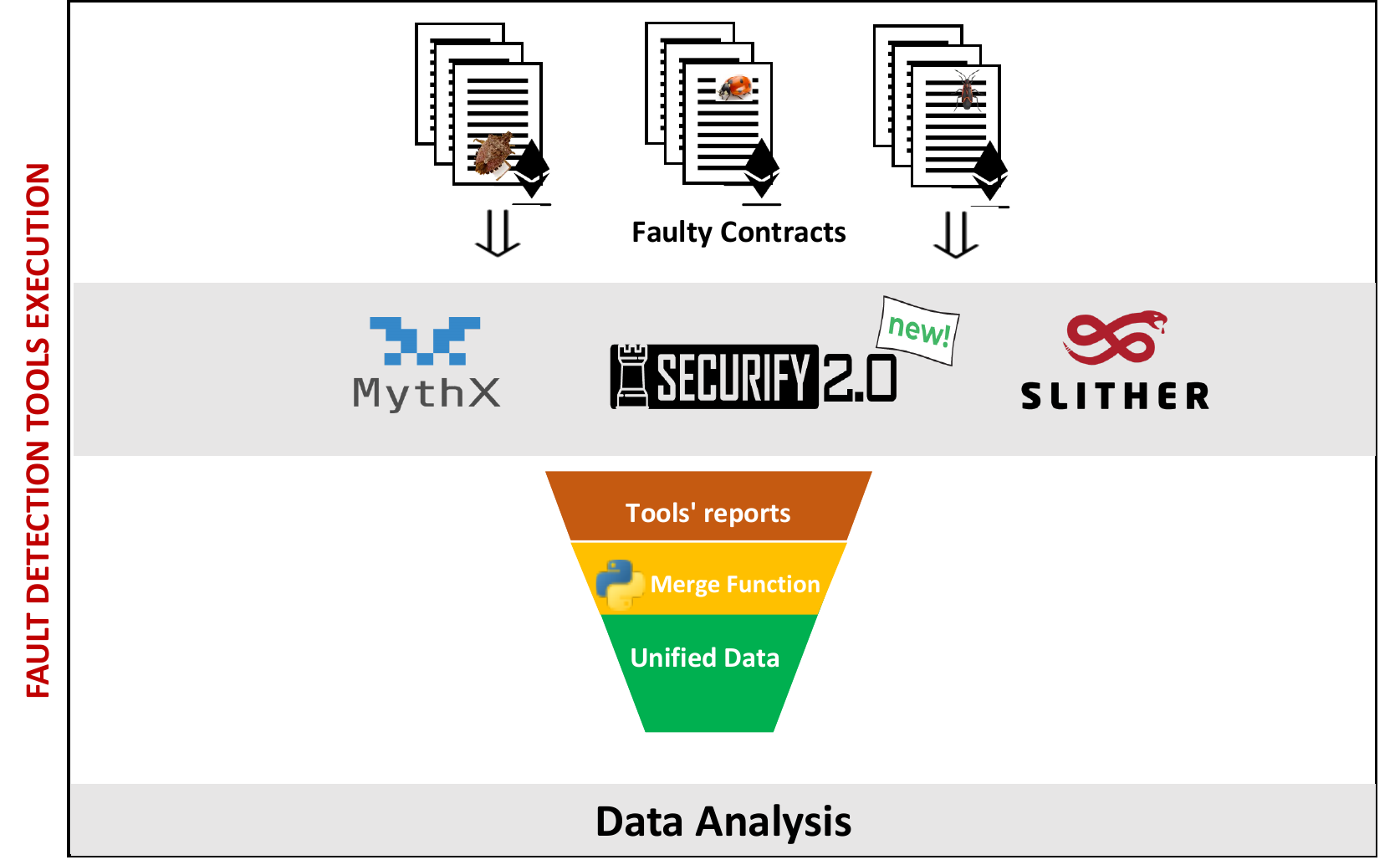}
    \caption{Approach for analyzing the effectiveness of verification tools.}
    \label{fig:experiment-2-approach}
\end{figure}

\begin{enumerate}
    \item Selection of smart contract verification tools;
    \item Execution of the tools against faulty smart contracts (generated by fault injection);
    \item Results analysis, based on a set of metrics of interest.
\end{enumerate}

\textit{Step 1)} involves the \textbf{selection of smart contract verification tools}. We aimed at popular tools, actively maintained, and of different operational nature. Namely, we selected an abstract interpretation tool (Securify2 version 0.0.1), a static analysis tool (Slither version 0.8.0), and a tool that uses symbolic execution (Mythril version 0.22.19). In the perspective of our approach, this is a variable set of tools and, at this point, other tools could be used (e.g., Zeus \cite{kalra_zeus_2018}, Oyente \cite{luu_making_2016}, Smartest \cite{Sunbeom2021}, 
Smartian \cite{Choi2021a}).The specific selection of tools will depend on various factors, such as the available time to run the tools and to analyze results (e.g., some tools require more time to execute, other tools have high false positive rates), computational resources, and the overall requirements of the user executing the approach.

In \textit{Step 2)}, we \textbf{execute the tools} against the generated set of 15,494 faulty smart contracts, collect their output and then store and process the results produced by the tools, mapping the detected vulnerabilities to the analyzed contracts. The tools are run using their default parameters, with no particular configuration towards specific types of faults.

In \textit{Step 3)}, \textbf{we analyze the results} but right before that, the tools' output reports are converted into unified data format through a \textit{merge function} we implemented. This way, no changes in the analysis process are required for a new tool.
While analyzing the results, all cases of potential true-positives (i.e., software faults signaled by the tool that do exist) are manually verified to check if the signaled defect really is present in the contract (also as a way of understanding if the fault injector is correctly injecting the faults). We focus on evaluating the tools’ overall effectiveness in detecting the injected faults, which should be present in all contracts under analysis. Other potential faults (i.e., previously unknown vulnerabilities) are out of the scope of this work.

\subsection{Study 3 - Impact of Elusive Faults}

This final study focuses on the outcomes of the previous studies and analyses the consequences of the faults that elude the verification tools. The analysis is essentially carried out to understand the distribution of faulty contracts (not detected by any of the tools) per fault type; the prevalence of the different types of failures associated with the different types of faults; and, finally, the study focuses on the faults that generate the most severe failures.


\section{Results and Discussion}
\label{sec:results}
\begin{figure*}[th]
\centering
\includegraphics[width=1.0\linewidth]{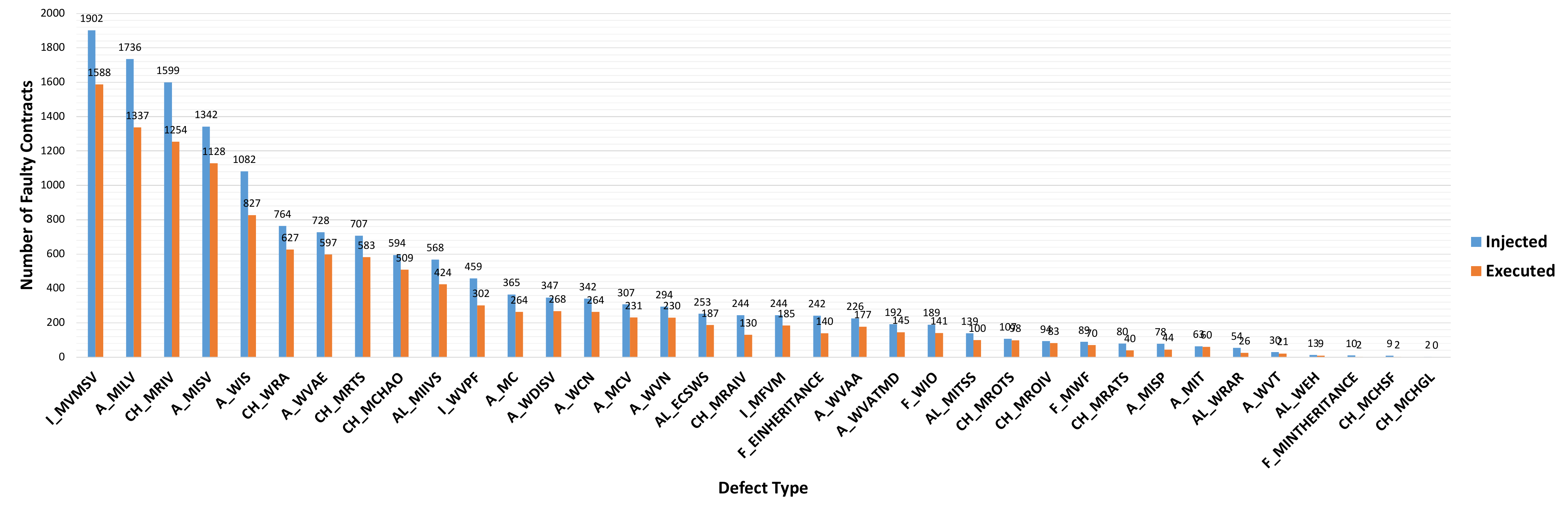}
\caption{Distribution of the generated faulty smart contracts per defect type.}
\label{fig:distribution}
\end{figure*}

This section discusses the results obtained during our experimental evaluation.
All the experiments were executed on 4 virtual machines with 16 CPU, 16 memory, using Ubuntu 18.04.5 LTS.
After running the fault injection process, we were able to generate \textbf{at least one} faulty contract (out of 400 smart contracts) for each of the 36 different types of faults, ending up with a total of 15,494 (>= 400 * 36 as it is possible to inject a single fault in more than one location in the code of a certain contract) faulty smart contracts. Figure \ref{fig:distribution} overviews the distribution of the generated faulty contracts per defect type (blue bars). 

As we can see in Figure \ref{fig:distribution}, some faults lead to higher numbers of faulty contracts, such as \textit{Missing visibility modifier of state variables (I\_MVMSV)} (1902 times), \textit{Missing initialization of Local Variable (A\_MILV)} (1736 times), and \textit{Missing require on input values (CH\_MRIV)} (1599 times). These numbers are not directly reflect their frequency in the real world, but are simply related with the number of possible locations in each of the original contract's code that met the conditions for the injection. On the opposite side we find a few faults that appear rarely, such as, \textit{Missing Check on Gas Limit (CH\_MCHGL)} (2 times), or \textit{Missing Check on Suicide Functionality (CH\_MCHSF)} (9 times). Notice that, although a fault may be infrequent (i.e., low probability of occurrence) the associated risk may be high, which means that tools should not disregard such cases. 

Figure \ref{fig:distribution} also shows, in the orange bars, the number of faulty smart contracts that were actually executed for each defect type. As the workload generation tool (described in section \ref{sec:approach}) is currently unable to fully match types and number of parameters necessary for invoking all transactions in all 15,494 contracts, the number of executed contracts is less than the total number of contracts. Still, we were able to run 83\% (12093 out of 15,494) of all generated faulty contracts, encompassing all 36 types of faults.

\subsection{Results of Study 1 - Faults' Impact}

\begin{figure}[!ht]
    \centering
    \includegraphics[scale=0.55]{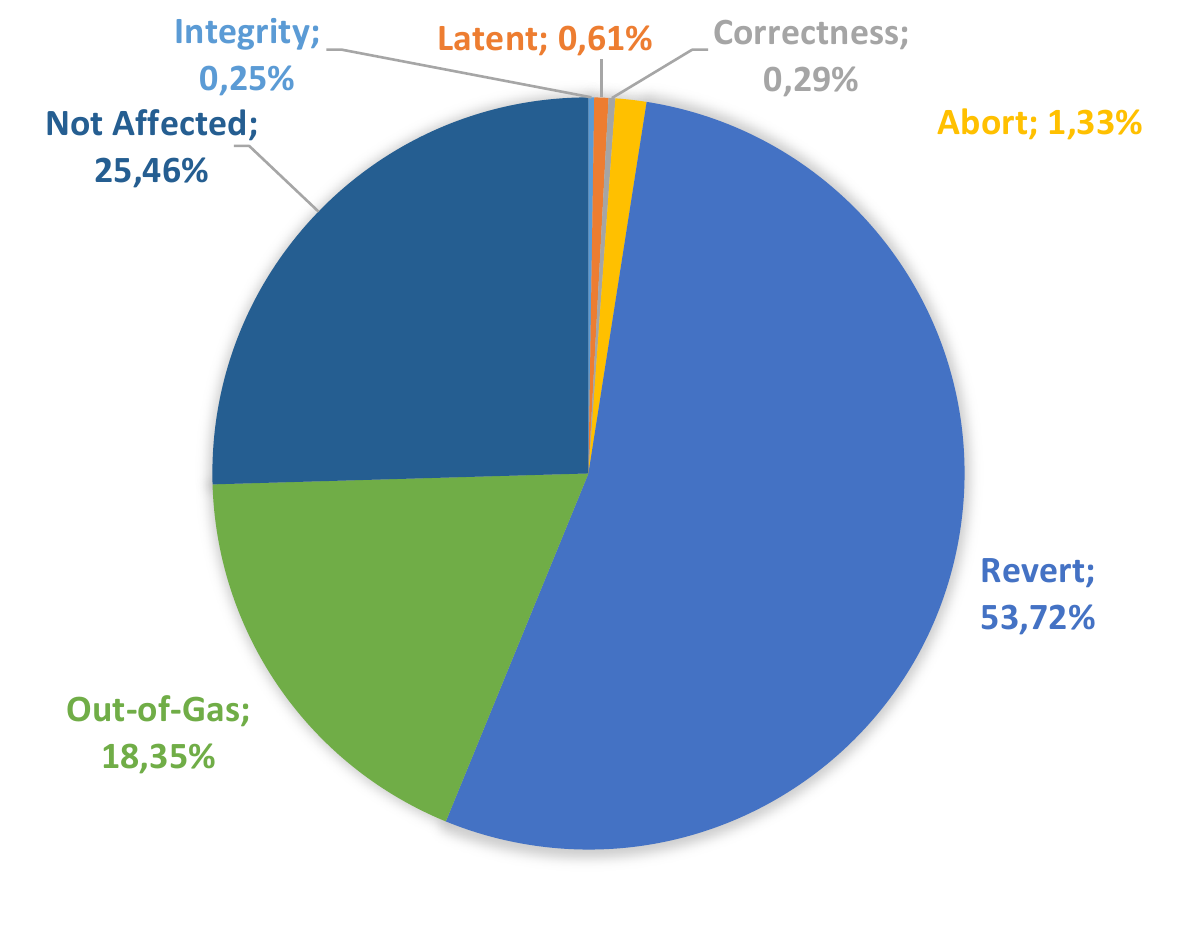}
    \caption{Overall view of faults' impact.}
    \label{fig:impact_overal}
\end{figure}

We ran the generated workload over the faulty smart contracts, which resulted in the execution of a total of $10,925,749$ transactions of which $2,782,063$ (about $25.46$\%) were executed successfully and no effect was observed. The rest of the transactions were affected by the injected fault having resulted in a failure. Figure \ref{fig:impact_overal} shows an overview of the distribution of results.

Most of the failures triggered are of type \textit{Revert Failure} (about 53.72\% of all transactions) followed by \textit{Out-of-Gas Failure} (about 18.35\% of all transactions). The rest of the failures, which are the most critical ones (as they influence on gas consumption and correctness of results and ledger), compose less than 2.5\% of the cases. 

\begin{figure*}[h]
    \centering
    \includegraphics[width=1.0\textwidth]{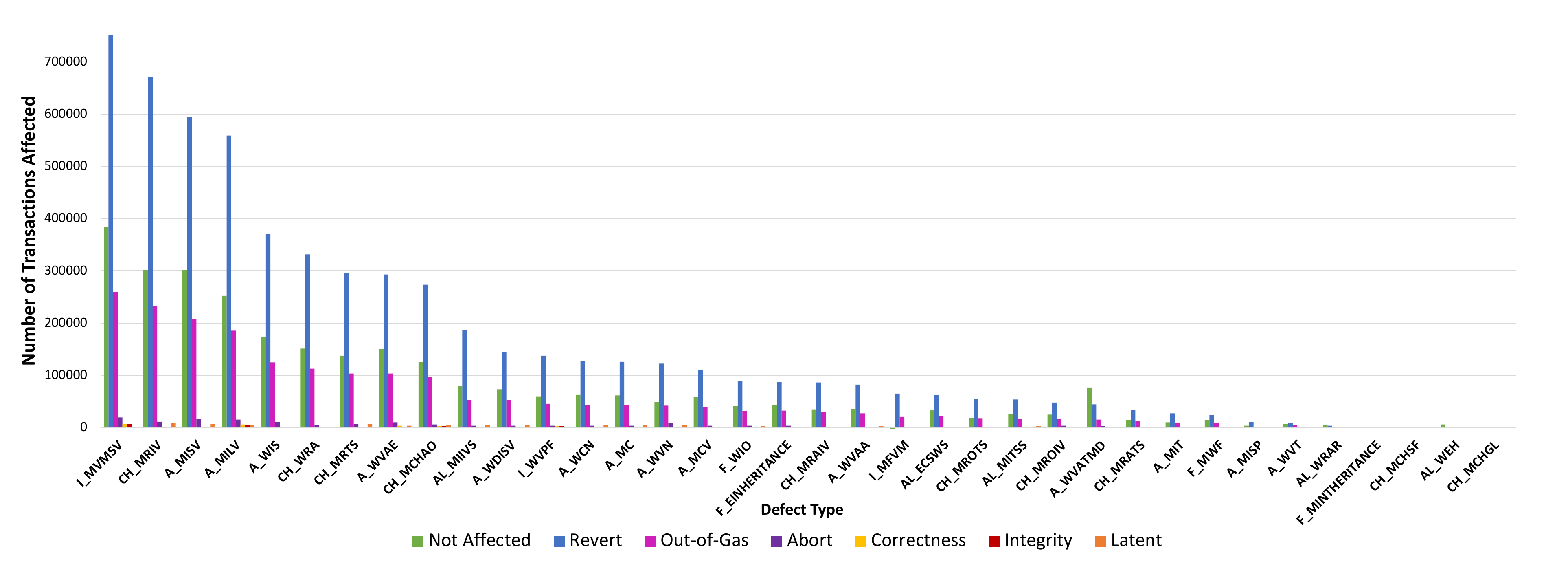}
    \caption{Faults' impact per defect type.}
    \label{fig:impact_per_defect}
\end{figure*}

Figure \ref{fig:impact_per_defect} shows the detailed impact results per defect type. Notice that drilling down to the fault type, the relative prevalence of the different types of failures is generally the same across all types of faults.

\begin{figure}[h]
    \centering
    \includegraphics[scale=0.65]{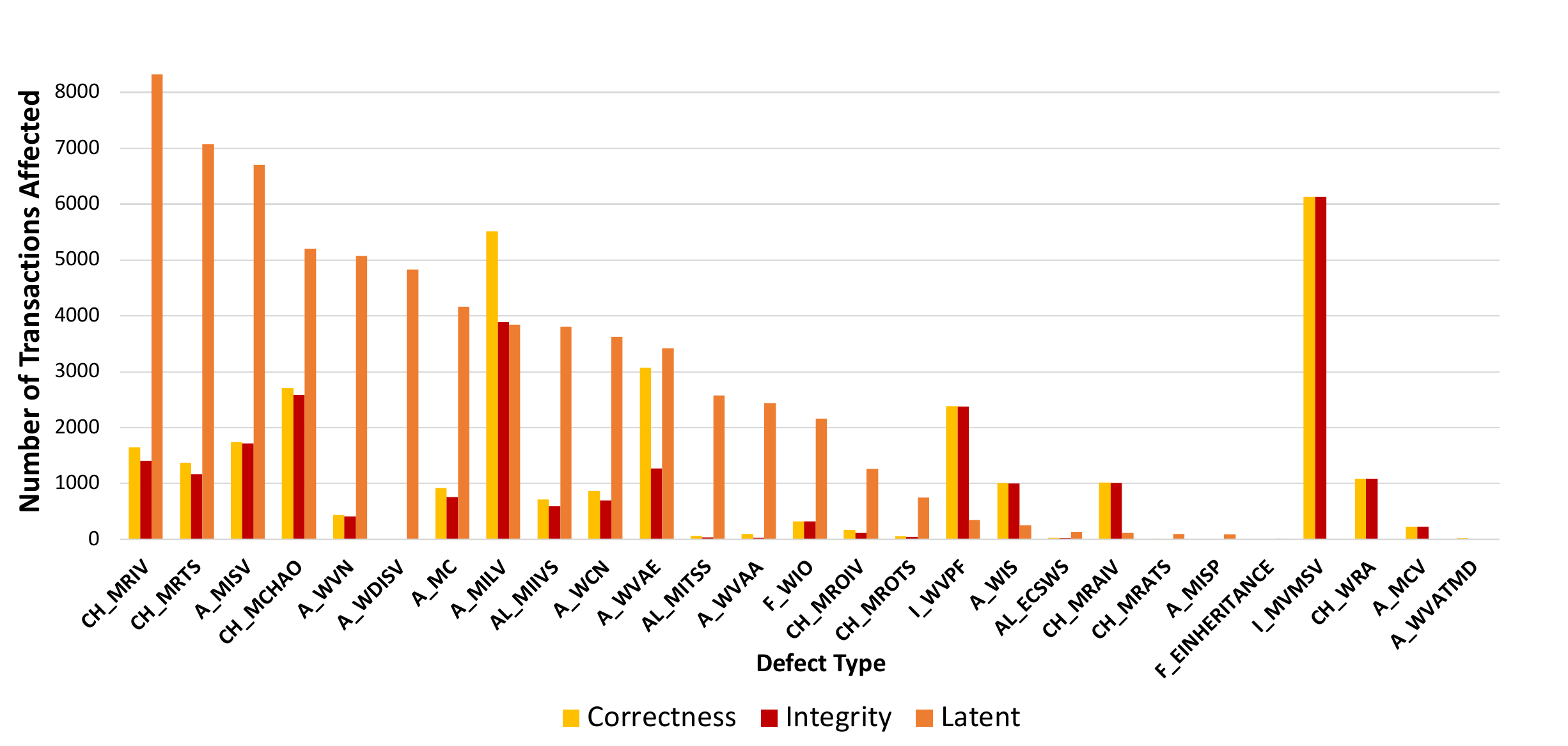}
    \caption{Types of faults triggering severe failures.}
    \label{fig:impact_per_defect_critical}
\end{figure}

Figure \ref{fig:impact_per_defect_critical} shows the detailed results of the fault types that caused severe failures namely \textit{Correctness Failure}, \textit{Integrity Failure}, and \textit{Latent Integrity Failure}. Of all 36 types of faults, only 9 of them did not cause any of these failures (for instance, \textit{CH\_MCHGL} and \textit{CH\_MCHSF} are two of these 9 cases). Note that all of these 9 fault types, with the exception of \textit{I\_MFVM}, are the least frequent in our faulty smart contracts list (refer to Figure \ref{fig:distribution}). 

The results depicted in Figure \ref{fig:impact_per_defect_critical} show that there are still many cases in which most of the defect types injected cause correctness, integrity, and especially latent integrity failures.
As shown, \textit{Missing require on input variables (CH\_MRIV)} causes most of \textit{Latent} failures and \textit{Missing visibility modifier of state variables (I\_MVMSV)} causes most of \textit{Integrity} and \textit{Correctness} failures.
Among all, \textit{Missing if construct on transaction sender plus statements (AL\_MITSS)} and \textit{Wrong variable name (A\_WVN)}, respectively with 3.12\% and 2.75\%, have a higher ratio (total number of failures divided by the total number of transactions executed per defect type) of \textit{Latent Integrity Failure}.

\begin{figure}[h]
    \centering
    \includegraphics[width=0.5\textwidth]{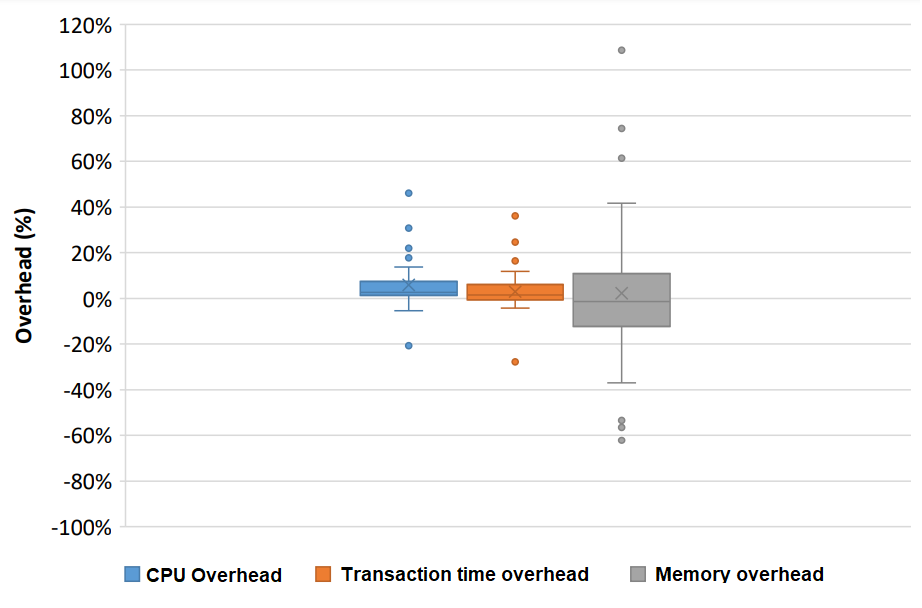}
    \caption{Overhead caused by faults on CPU, Memory and Execution time.}
    \label{fig:overhead}
\end{figure}

We have also calculated the runtime overhead (performance degradation) caused by injected faults on faulty contracts compared to fault-free runs in terms of CPU usage, Memory Usage, and Transaction time. An overview of the results is presented in Figure \ref{fig:overhead}, in which we can see the distribution of the three types of overhead values for all transactions. In general, the injected faults lead to some overhead on all three metrics. There are some cases where the overhead is high, namely \textit{Wrong arithmetic expression used in assignment (A\_WVAE)} on CPU usage, \textit{Wrong variable type (A\_WVT)} on transaction time, and \textit{Wrong value assignment with too many digits (A\_WVATMD)} on memory usage. It is also worthwhile mentioning that some faults are associated with negative overhead values since they lead to \textit{abort} or \textit{revert} of transactions. 

\subsection{Results of Study 2 - Effectiveness of Verification Tools}

Table \ref{tab:odc_vs_tools} shows which faults the three tools (i.e., Securify, Slither, and Mythril) announce they are able to detect. It shows the original names used by the tools and maps them to our fault model.


\begin{table}[h]
\centering
\caption{Mapping of our fault model to the defects detected by the tools.}
\label{tab:odc_vs_tools}
\begin{tabular}{|l|l|l|l|}
\hline
\textbf{Fault}  & \textbf{Securify}                         & \textbf{Slither}                             & \textbf{Mythril} \\ \hline
A\_MC           & CallToDefaultConstructor?                 & void-cst                                     & SWC-118          \\ \hline
A\_MCV          & -                                         & -                                            & SWC-102          \\ \hline
A\_MILV         & {\color[HTML]{24292F} UninitializedLocal} & {\color[HTML]{24292F} uninitialized-local}   & SWC-109          \\ \hline
A\_MISP         & UninitializedStorage                      & {\color[HTML]{24292F} uninitialized-storage} & SWC-109          \\ \hline
A\_MISV         & UninitializedStateVariable                & uninitialized-state                          & SWC-109          \\ \hline
A\_WCN          & CallToDefaultConstructor?                 & void-cst                                     & SWC-118          \\ \hline
A\_WDISV        & {\color[HTML]{24292F} ConstableStates}    & constable-states                             & -                \\ \hline
A\_WIS          & -                                         & storage-array                                & SWC-101          \\ \hline
A\_MIT          & -                                         & divide-before-multiply                       & SWC-101          \\ \hline
A\_WVAA         & -                                         & {\color[HTML]{24292F} missing-zero-check}    & -                \\ \hline
A\_WVAE         & -                                         & -                                            & -                \\ \hline
A\_WVATMD       & {\color[HTML]{24292F} TooManyDigits}      & {\color[HTML]{24292F} too-many-digits}       & SWC-101          \\ \hline
A\_WVN          & ShadowedStateVariable                     & shadowing-state                              & SWC-119          \\ \hline
A\_WVT          & -                                         & controlled-array-length                      & -                \\ \hline
AL\_ECSWS       & CallInLoop                                & calls-loop                                   & SWC-104          \\ \hline
AL\_MIIVS       & -                                         & -                                            & SWC-123          \\ \hline
AL\_MITSS       & UnrestrictedEtherFlow                     & unchecked-send                               & SWC-105          \\ \hline
AL\_WEH         & UnhandledException                        & unchecked-lowlevel                           & -                \\ \hline
AL\_WRAR        & -                                         & {\color[HTML]{24292F} assert-state-change}   & SWC-110          \\ \hline
CH\_MCHAO       & -                                         & -                                            & SWC-123          \\ \hline
CH\_MCHGL       & -                                         & {\color[HTML]{24292F} costly-loop}           & SWC-128          \\ \hline
CH\_MCHSF       & UnrestrictedSelfdestruct                  & suicidal                                     & SWC-106          \\ \hline
CH\_MRAIV       & -                                         & -                                            & SWC-123          \\ \hline
CH\_MRATS       & -                                         & -                                            & SWC-123          \\ \hline
CH\_MRIV        & -                                         & -                                            & SWC-123          \\ \hline
CH\_MROIV       & -                                         & -                                            & SWC-123          \\ \hline
CH\_MROTS       & -                                         & -                                            & SWC-123          \\ \hline
CH\_MRTS        & -                                         & -                                            & SWC-123          \\ \hline
CH\_WRA         & TxOrigin                                  & tx-origin                                    & SWC-115          \\ \hline
F\_EINHERITANCE & -                                         & missing-inheritance                          & SWC-125          \\ \hline
F\_MINHERITANCE & -                                         & missing-inheritance                          & SWC-125          \\ \hline
F\_MWF          & LockedEther                               & locked-ether                                 & -                \\ \hline
F\_WIO          & -                                         & missing-inheritance                          & SWC-125          \\ \hline
I\_MFVM         & ExternalFunctions                         & external-function                            & -                \\ \hline
I\_MVMSV        & StateVariablesDefaultVisibility           & -                                            & SWC-108          \\ \hline
I\_WVPF         & -                                         & constant-function-asm                        & -                \\ \hline
\end{tabular}
\end{table}

\begin{figure}[h]
\centering
\includegraphics[scale=0.5]{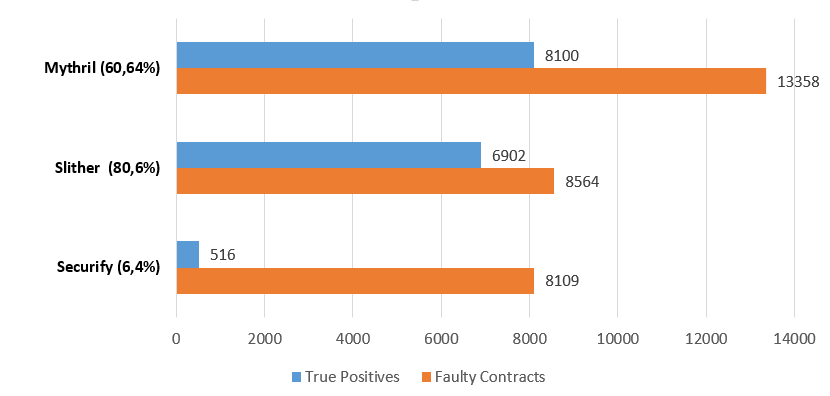}
\caption{Detection accuracy per tool.}
\label{fig:tools_injected_vs_projected1}
\end{figure}

Figure \ref{fig:tools_injected_vs_projected1} shows an overview of the detection accuracy of each of the three tools used. In particular, it shows, per tool, the total number of faulty contracts generated (considering only the types of faults each tool was designed to detected) and the total number of contracts in which the tools signaled the presence of a problem in the injection location (i.e., the true positives).

As we can see in Figure \ref{fig:tools_injected_vs_projected1}, Slither is more effective in detecting the injected defects (detects defects in about 81\% of the contracts) and is followed by Mythril with about 61\% of detection accuracy. Securify shows clearly lower values of detection accuracy reaching only about 6\%. We emphasize that these accuracy numbers use the announced capabilities of each of the tools as reference.

It is important to mention that although Slither seems to be a more effective verification tool among the three tools evaluated in this study, the number of alerts generated by Slither is also much higher than the number of alerts generated by other tools. During the experiments and considering just the faulty contracts holding faults that each of the tools were designed to detect, Securify generated a total of $7382$ alerts, of which $516$ were indeed correct alerts (i.e., $6.99$\% of the alerts represented true positives). Mythril generated 55090 alerts, of which 8100 ended up being correct alerts ($14.70$\%). Slither generated 397236 alerts, of which only 6902 were correct alerts ($1.74$\%).

\begin{figure}[h]
\centering
\includegraphics[scale=0.45]{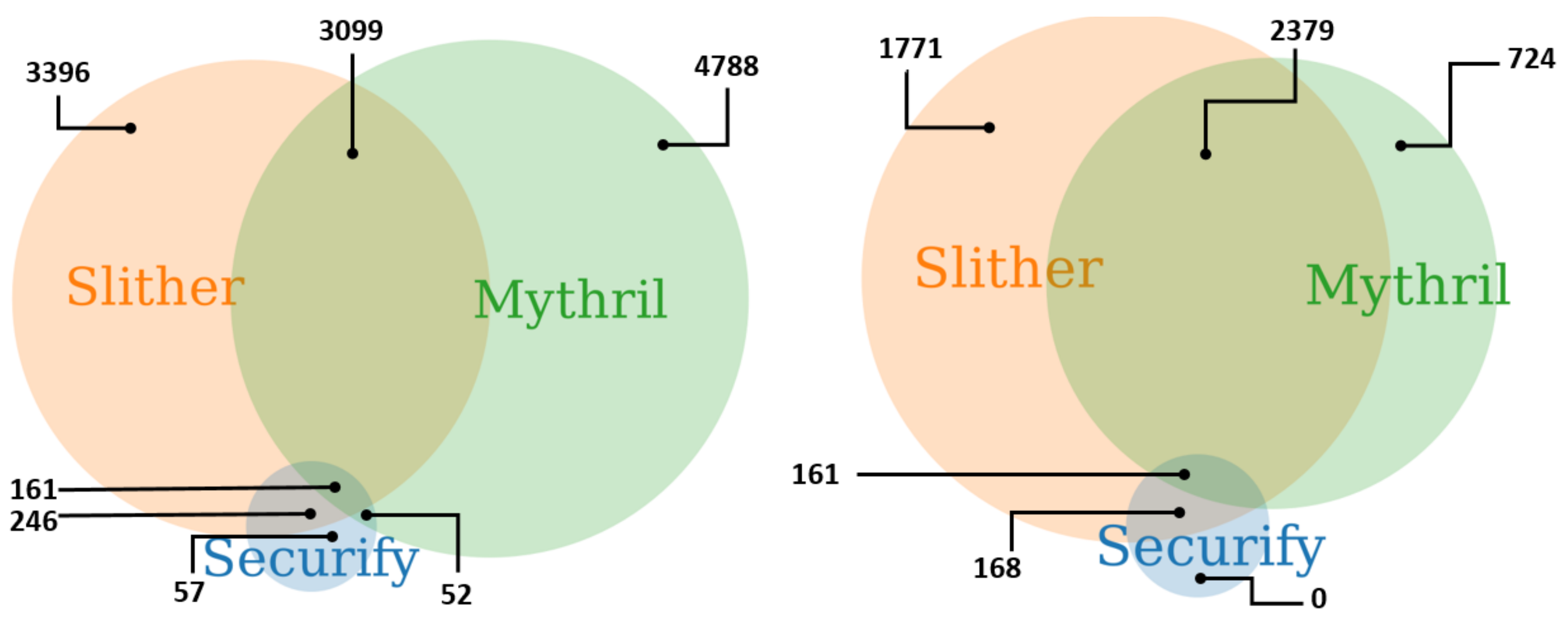}
\caption{Venn diagram of detected defects by the Tools.}
\label{fig:detected_venn}
\end{figure}

Figure \ref{fig:detected_venn} shows how differently the verification tools performed in detecting the faulty contracts. Figure \ref{fig:detected_venn}.a) at the left-hand side considers all faults which the tools were designed to detect (including faults that only one or two of the tools should detect). Figure \ref{fig:detected_venn}.b) considers only the set of faults that are common to the three tools, i.e., that all three tools announce being able to detect.

As we can see in Figure \ref{fig:detected_venn}.a), only 161 (about 1.4\%) faulty contracts out of 11799 are signaled correctly by all three tools. We can also see that 3099 faulty contracts (about 26.3\%) are detected by both Slither and Mythril. The rest of the faulty contracts are detected either by Slither or by Mythril, with the advantage being on the side of Mythril. Although Securify has low detection effectiveness it can actually signal faults in 57 contracts that neither of the remaining tools are able to. This clearly shows the tools complementarity in fault detection. In Figure \ref{fig:detected_venn}. b) we again observe the complementarity of the tools, although we now see that Slither actually captures most of the faults that Mythril detected (in this scenario were we reduced the faults to the set that is common to the three tools). We also see that Securify does not bring further detection value in this scenario.

\begin{figure*}[h]
\centering
\includegraphics[width=1.0\linewidth]{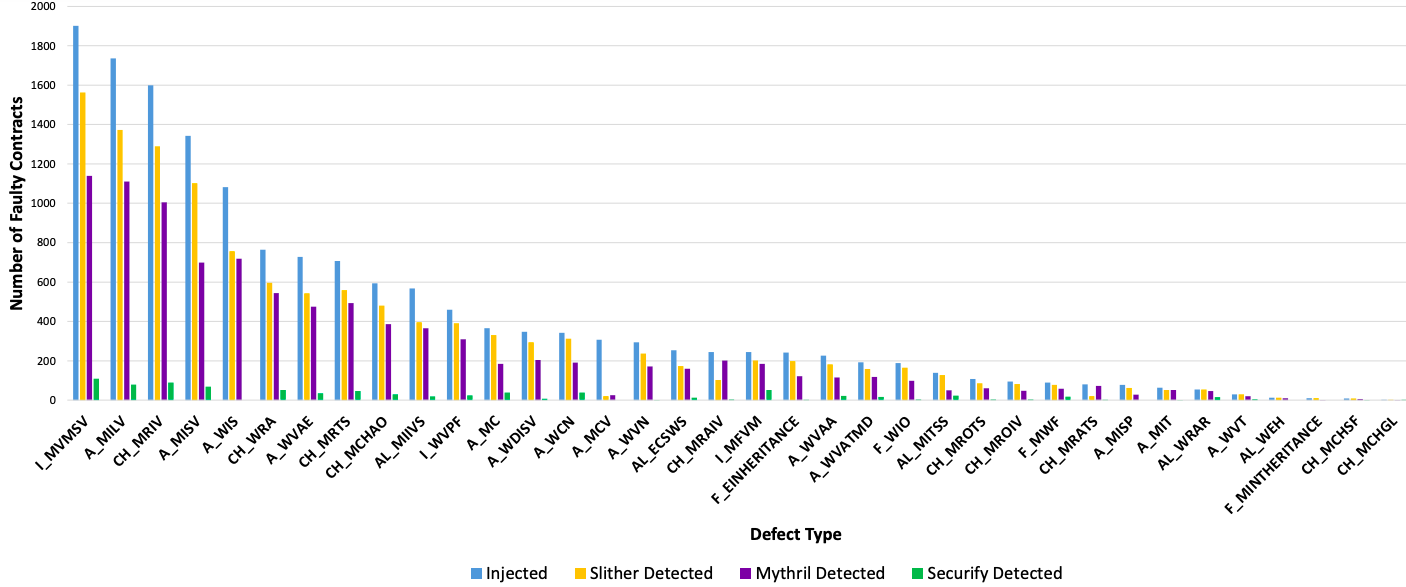}
\caption{Defects detected by the verification tools.}
\label{fig:detected}
\end{figure*}

We now go through a more detailed view of the tools capabilities per each of the faults. Figure \ref{fig:detected} presents, per type of fault, the number of faulty contracts generated and the corresponding number of contracts in which the tools signaled the presence of a problem in the injection location (i.e., the tools detection accuracy). 

As we can see in Figure \ref{fig:detected}, the pattern of detection seems to be similar for all fault types, with exception of a few cases, namely \textit{A\_MCV}, \textit{CH\_MRAIV}, and \textit{CH\_MRATS} in which Mythril was able to detect more faulty contracts. In the case of \textit{Missing Compiler Version (A\_MCV)}, most of the faulty contracts have remained undetected. In contrast, defect types of \textit{A\_WIS}, \textit{A\_WVT}, \textit{AL\_WRAR}, \textit{CH\_MCHGL}, \textit{CH\_MCHSF}, \textit{F\_MINTHERITANCE} are totally detected by one or more tools.

\subsection{Results of Study 3 - Impact of Elusive Faults}
\label{sec:study3}

\begin{figure}[th]
\centering
\includegraphics[width=0.8\textwidth]{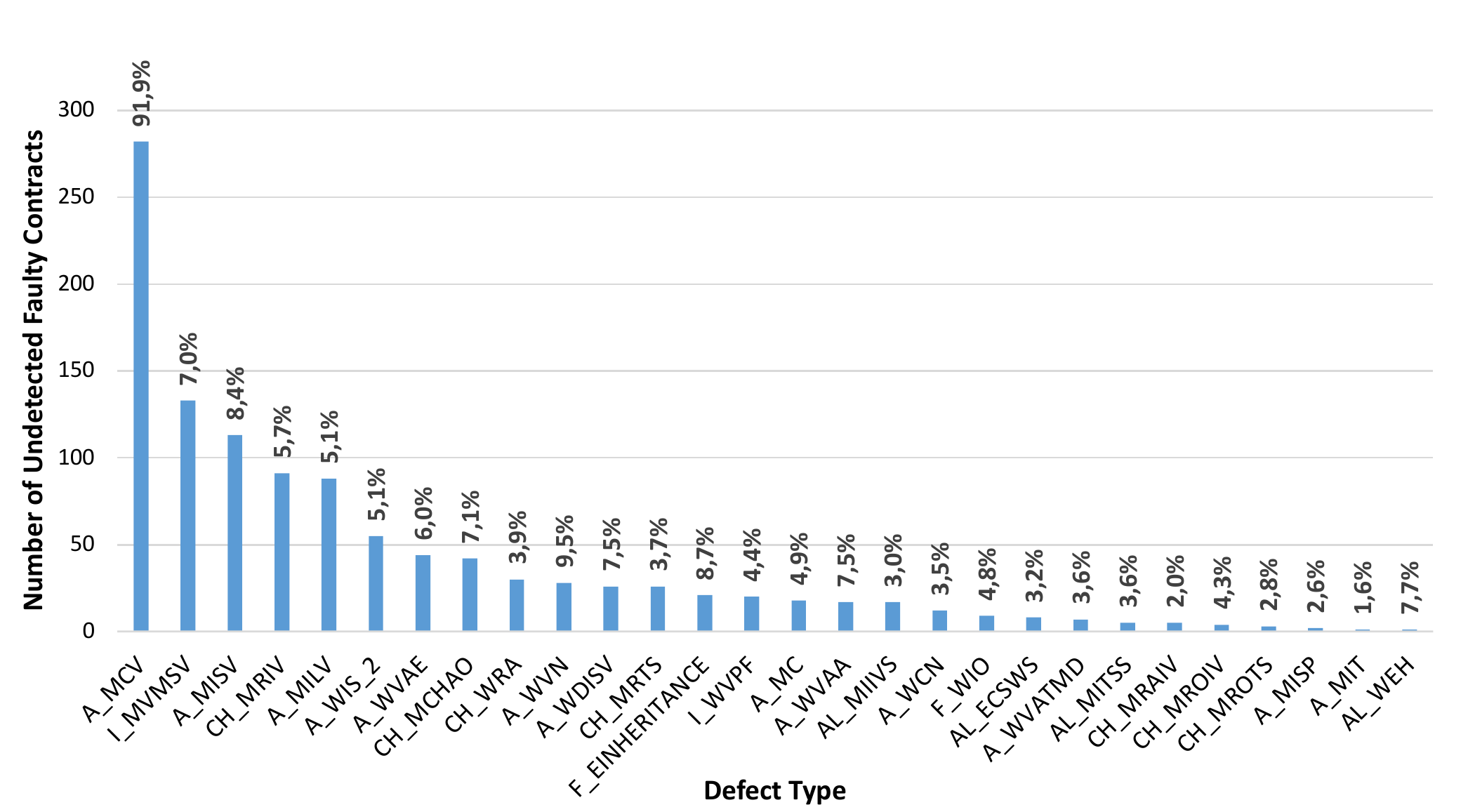}
\caption{Faulty Contracts not detected by any of the tools.}
\label{fig:notdetected}
\end{figure}

This section focuses on the faults that escaped detection by the verification tools and analyses their impact. Figure \ref{fig:notdetected} presents the number of faulty contracts that are not detected by any of the verification tools, per defect type. The labels on top of each bar reflect the percentage of undetected faulty contracts of a certain type of fault, considering the total number of faulty contracts generated for that particular fault. In total, about 9\% of the contracts (1395 out of 15494) remained undetected by all tools. 
The defect type \textit{Missing Compiler Version (A\_MCV)} should be simple to detect (it can be done by a simple check at the beginning of the contract), however, the tools generally fail to detect it in most cases (91.9\%). 
In the case of the other defect types, tools tend to perform better and, the injected fault is detected by at least one of the tools in at least every 9 out of 10 faulty contracts. Still, the different code locations where the fault was injected affects the detection capabilities of the tool. 

\begin{figure}[!ht]
    \centering
    \includegraphics[scale=0.5]{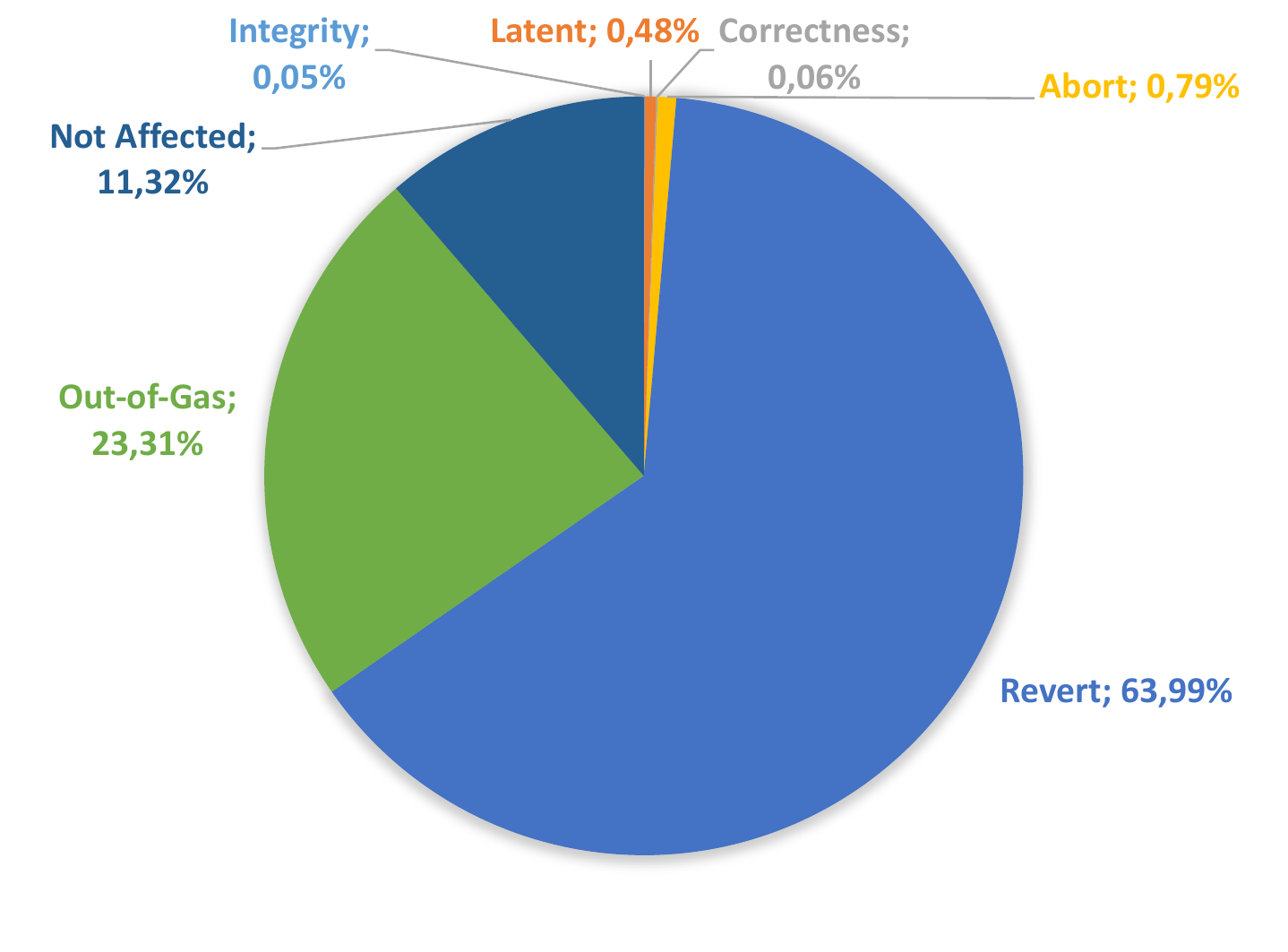}
    \caption{Overall view of undetected faults' impact.}
    \label{fig:impact_overal_undetected}
\end{figure}

To understand the impact of the undetected defects, we present a summary of the results obtained from executing the transactions of the undetected faulty contracts in Figure \ref{fig:impact_overal_undetected}. Again, we observe that most of the failures belong to \textit{Revert Failure} and \textit{Out-of-gas Failure} followed by \textit{Abort Failure}. When compared to the distribution of failures in all transactions, presented in Figure \ref{fig:impact_overal}, the percentage of not affected transactions decreased among the elusive faults leading to a higher percentage of \textit{Revert} and \textit{Out-of-gas} failures. The percentage of the other failures slightly decreased as well. 

\begin{figure}[t]
\centering
\includegraphics[width=1.0\textwidth]{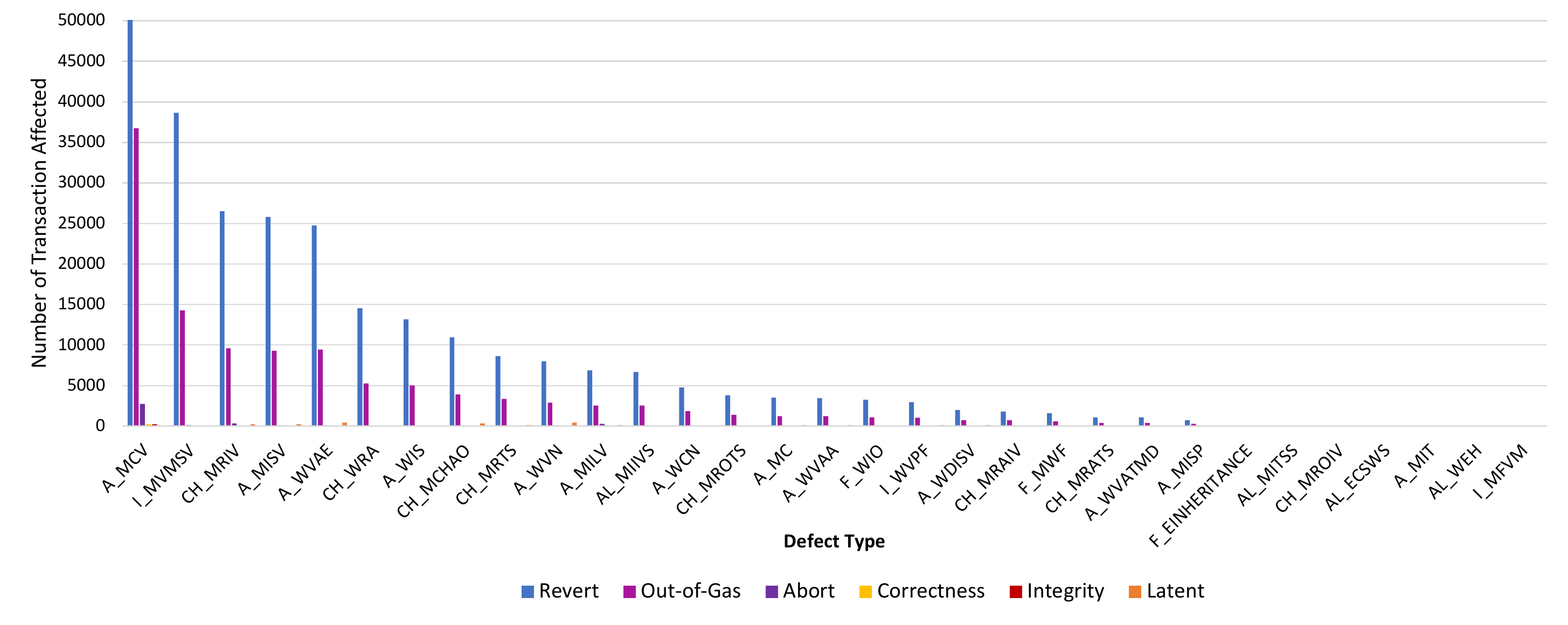}
\caption{Faults' impact for undetected defects.}
\label{fig:impact_undetected}
\end{figure}

A more detailed view of the results is presented in Figure \ref{fig:impact_undetected}. 
 Among undetected defects, \textit{A\_MCV} is causing most of the failures. In contrast, undetected defects types of \textit{I\_MFVM}, \textit{AL\_WEH}, \textit{A\_WVATMD\_2}, \textit{A\_MIT}, \textit{AL\_ECSWS}, \textit{CH\_MROIV}, \textit{F\_INHERITANCE} are not causing any failure. 

\begin{figure}[t]
\centering
\includegraphics[scale=0.85]{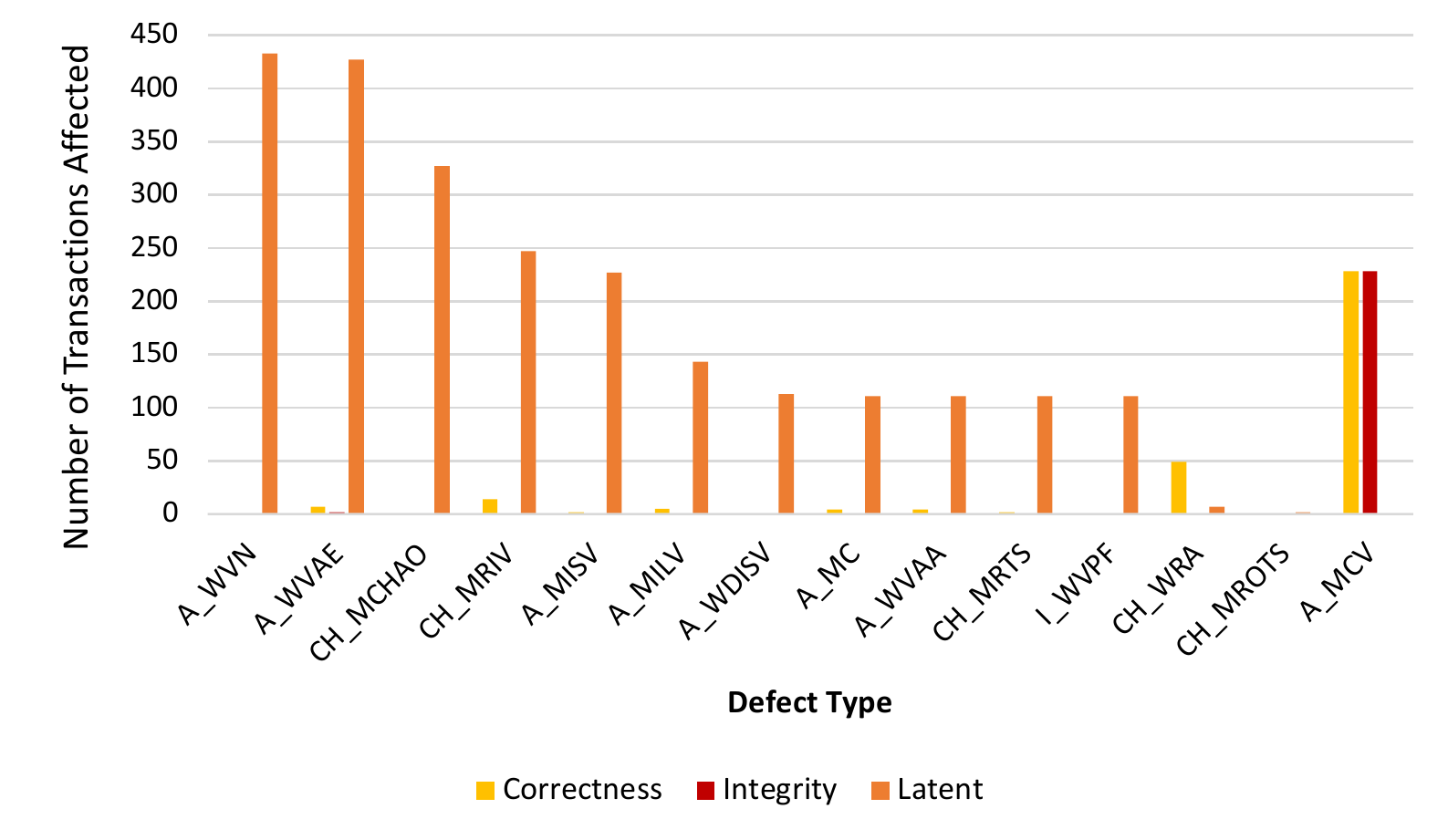}
\caption{Faults' critical impact for undetected defects.}
\label{fig:impact_undetected_critical}
\end{figure}

Figure \ref{fig:impact_undetected_critical} drills down to the undetected faults that caused severe failures namely \textit{Correctness Failure}, \textit{Integrity Failure}, and \textit{Latent Integrity Failure}. The results show that even after using the whole set of verification tools, residual faults are indeed left behind and leading to severe issues, from the blockchain point of view. The results also show that most of these severe defects are either of type \textit{Assignment} or \textit{Checking}.

Among all types of faults presented in Figure \ref{fig:impact_undetected_critical}, \textit{A\_MCV}, \textit{CH\_MROTS}, \textit{CH\_WRA} are less severe as they are not causing any or a just a few latent failures. In contracts \textit{A\_WVN} and \textit{A\_WVAE} can be assumed as the most severe issues we may find in smart contracts, as both are causing latent failures in most cases. The former remained undetected for about 9.5\% of the time, and the latter remained undetected for about 6.0\% of the time (refer to Figure \ref{fig:notdetected}). 

\subsection{Main Findings}
This section highlights the main findings of our experimental evaluation, as follows:

\begin{itemize}
    \item As a general observation related with the fault injection process, we found out that a few type of faults are connected to higher likelihood of injection, namely \textit{Missing visibility modifier of state variables (I\_MVMSV)} (1902 times), \textit{Missing initialization of Local Variable (A\_MILV)} (1736 times), and \textit{Missing require on input values (CH\_MRIV)} (1599 times) lead to higher numbers of faulty contracts. This means that the conditions required to inject these faults are realized more frequently.
    
    \item No failures were observed in one fourth of the faulty contracts, while in about half of the faulty contracts  \textit{Revert failures} were detected, with \textit{Out-of-gas failures} being observed in nearly one fifth of the faulty contracts. These two types of failures are the most frequent ones observed in these experiments. 
    
    \item The faults associated with higher chances of injection (i.e., \textit{I\_MVMSV}, \textit{A\_MILV}, and \textit{CH\_MRIV}) are also the ones that lead to most of the \textit{Revert failures} and \textit{Out-of-gas failures} observed during the experiments. 
    
    \item Fault \textit{CH\_MRIV}, one of the most frequent, is responsible for most \textit{Latent failures},  which is the most severe failure mode. \textit{CH\_MRTS} and \textit{A\_MISV\_2} are not as frequent as \textit{CH\_MRIV} but they are also the cause of a visible number of cases of \textit{Latent failures}. 
    
    \item The effectiveness of smart contract verification tools is rather low, with results showing low numbers of true positives when compared to a large number of generated alerts. This confirms similar observations in related work. 
    Slither seems to be more effective in detecting the injected faults (it is able to detect defects in about 81\% of the faulty contracts), but it also generates a huge number of alerts (the detected defects compose only 1.74\% of all alerts generated). Mythril, which detects defects in about 61\% of faulty contracts, is an interesting option if we consider the number of alerts generated (the detected defects compose only 14.70\% of all alerts generated). Securify showed to be able to detect about 6.4\% of the faulty contracts.
    
    \item Mythril and Slither have clearly shown complementary capabilities, although they also jointly detected many of the faulty contracts. Securify was able to detect faults that the remaining tools could not capture, but at a very small scale. Thus, developing a tool that makes the use of the different techniques involved is a possible path towards better detection capabilities.
    
    \item Faulty contracts generated with \textit{A\_WIS}, \textit{A\_WVT}, \textit{AL\_WRAR}, \textit{CH\_MCHGL}, \textit{CH\_MCHSF}, \textit{F\_MINTHERITANCE} are totally detected by at least one of the tools. On the opposite side, the tools mostly fail to detect \textit{Missing Compiler Version (\textit{A\_MCV})}. 
    
    \item The faults generating the most severe failures either belong to \textit{Assignment} or \textit{Checking} defect types.
    
    \item The overall impact on CPU, memory and transaction time of the faults is relatively small (i.e., from 2 to 6\%), although there are concerning cases with some faults significantly exceeding the normal profile, in some cases duplicating the reference values (e.g., memory overhead).
    
    \item In what concerns the \textit{elusive faults} (see Section \ref{sec:study3}), nearly three quarters of the types of faults (28 out of 36) have escaped detection and are associated with severe failures (i.e., correctness, integrity, latent).
    
    \item \textit{A\_WVN}, \textit{A\_WVAE}, and \textit{CH\_MCHAO} are among the most severe issues, as they  jointly cause about 50\% of all latent failures in the transactions of undetected faulty contracts.

    \item The impact on CPU, memory usage, and transaction time is globally not significant, although the presence of \textit{A\_MC} and \textit{A\_WCN} in faulty contracts respectively leads to about 3 times more and 30 times more memory usage.
    
    \item Overall, focusing on the defect types identified as elusive during this work may allow for improving the detection capabilities of future verification tools. Also, a finer analysis per fault of the reasons of why a tool cannot detect the same fault in different code locations is crucial for detection improvement.

\end{itemize}

\subsection{Threats to Validity}

This section presents the threats to the validity of this work and discusses mitigation strategies. We start by mentioning that the \textit{fault model used} does not include all possible faults. For instance, we do not use reentrancy faults in this work, as well as other faults that are known to affect smart contracts. This may limit the evaluation of both impact and tools effectiveness and give a biased perception of the reality concerning impact and detection effectiveness. Anyway, the selected faults cannot be disregarded by detection tools nor their impact. Within this limitation, we did try to end up with at least one representative  example of each different type of fault. Using a more complete fault model and implementing a larger number of different faults will be pursued in future work.

The process for \textit{generating the workload} may not be the best option, considering that certain faults may only be triggered by very specific input sequences, which might shadow some interesting failures that could have occurred. Also, the characteristics of Solidity smart contracts may lead to calls that fail by specification (e.g., only some addresses have authorization and capabilities to perform transactions in the smart contracts). Nevertheless, we only analyze and compare transactions that are deemed successful in the base reference runs to make sure that the faulty reference runs indeed caused an impact.

The set of \textit{selected tools} is rather small and may not provide a proper view of smart contract verification tools. Also, depending on specific goals or constraints (e.g., available time and resources for executing the approach) other tools could be used; still, we selected tools that frequently appear in the literature. The \textit{analysis} is also mostly limited to measuring the true positive rate of the tools in detecting the presented vulnerabilities, which may not provide an accurate view of the tools' capabilities. Nevertheless, our goal is that our results allow improving verification tools, and the focus is on the injection of smart contract faults and generated faulty contracts, regardless of the tools that is then used for fault detection.

Finally, the whole combination of selected contracts with the implemented faults and selected tools may lead to a biased view of the faults that are indeed elusive. Still, we believe that our options were reasonable given the extension of the experiments and we highlight the presence of all three components in related work, supporting their representativeness.


\section{Conclusion}
\label{sec:conclusion}

In this work we carried out an experimental campaign to show the impact that realistic faults may have on the reliability of blockchain systems. We use fault detection tools to understand which of the faults may escape detection and if or how they lead the blockchain system to fail at runtime. In future work we intend to extend the implementation of the set of faults and use a larger and more diverse (in terms of operational profile) set of verification tools. Based on the foundations set in this paper, one of our future lines of research consists of the formal definition of a benchmark that allows assessing and comparing the effectiveness of vulnerability detection tools for smart contracts.




\bibliographystyle{plainnat}
{\footnotesize
\bibliography{library.bib}
}

\begin{IEEEbiography}[{\includegraphics[width=1in,height=1.25in,clip,keepaspectratio]{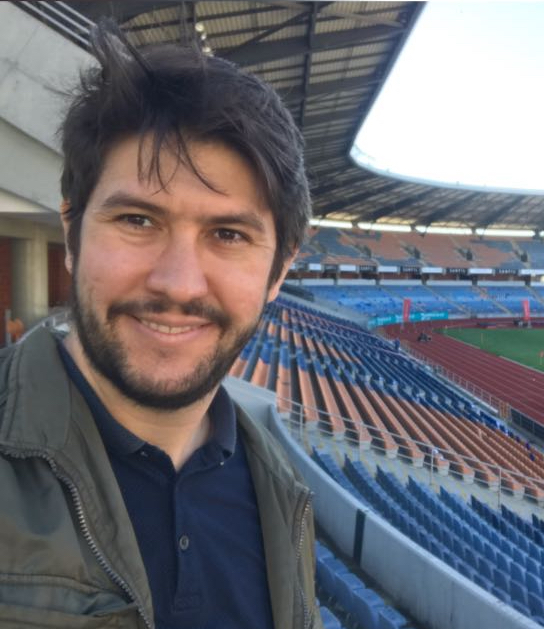}}] {Fernando Vidal} is a Ph.D. student at the University of Coimbra, Portugal, and his research interests are related to blockchain technology. Fernando has been publishing at international conferences, addressing some of his findings of blockchain technology,  such as vulnerabilities in smart contracts and revocation. In addition, Fernando was invited by the Advances in Science, Technology and Engineering Systems Journal (ASTESJ) magazine and IEEE Potentials to be one of the reviewers, of the blockchain submissions. Fernando has been applied his acquired knowledge, helping companies implement blockchain technology, through consulting.

\end{IEEEbiography}

\begin{IEEEbiography}[{\includegraphics[width=1in,height=1.25in,clip,keepaspectratio]{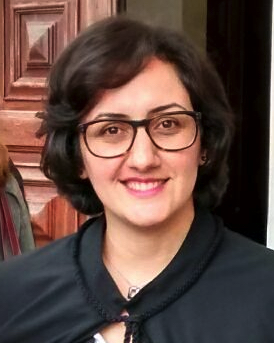}}]{Naghmeh Ivaki} received the PhD degree from the University of Coimbra, Portugal. Currently, she is an invited assistant professor and a full member of Software and Systems Engineering Group (SSE) of the Center for Informatics and Systems (CISUC), Department of Informatics Engineering, University of Coimbra. She specializes in the scientific field of Informatics Engineering, with particular focus on trustworthiness, security, safety, and dependability of computer systems. In her field of specialization, she has authored more than 30 peer-reviewed publications and participated in several national and international research projects. 
\end{IEEEbiography}

\begin{IEEEbiography}
[{\includegraphics[width=1in,height=1.25in,clip,keepaspectratio]{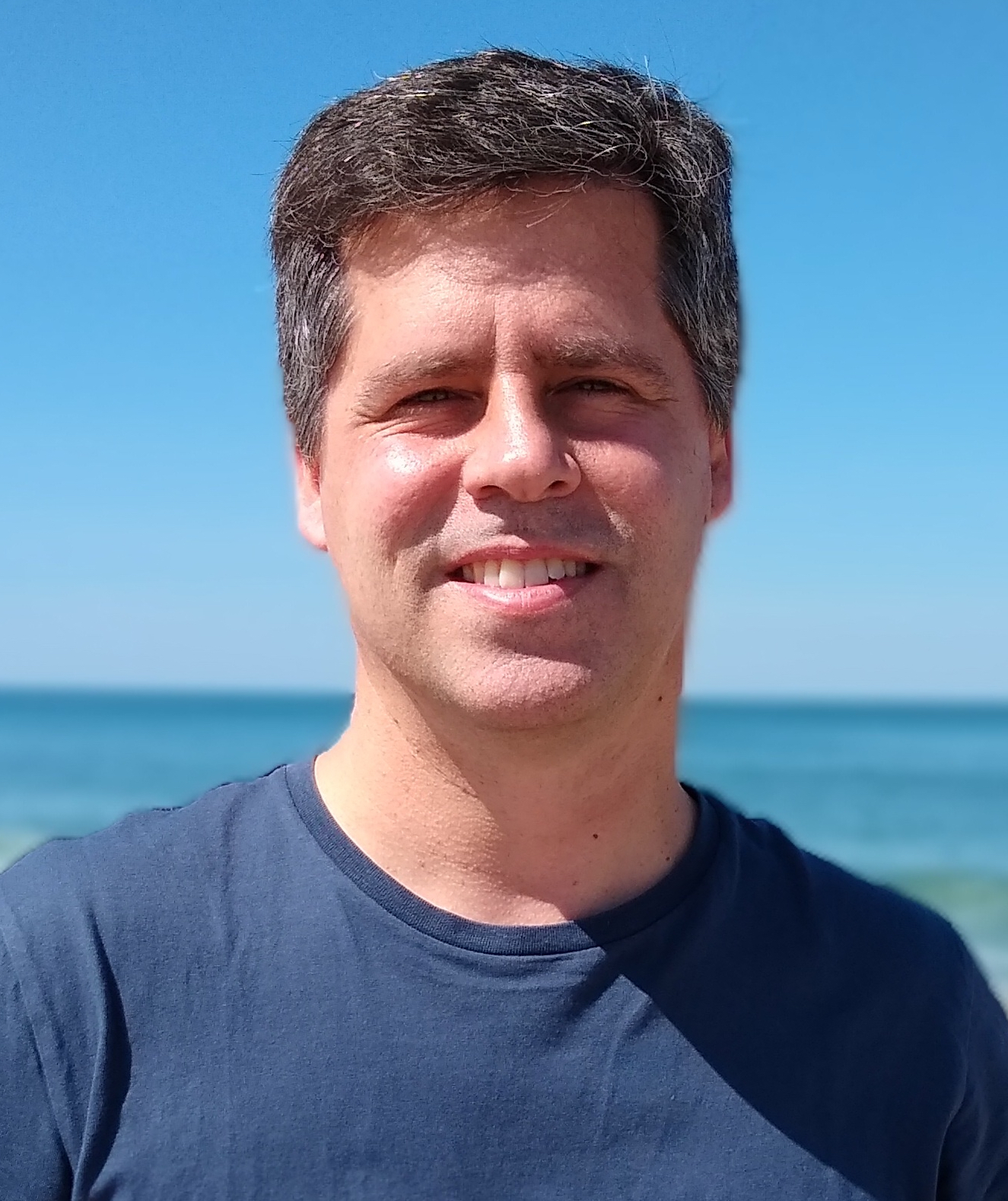}}]
{Nuno Laranjeiro} received the PhD degree from the University of Coimbra, Portugal, where he currently is an Assistant Professor. His research focuses on dependable and secure software services and he currently leads the Software and Systems Engineering group at the Centre for Informatics and Systems of the University of Coimbra (CISUC). His research interests include experimental dependability evaluation, fault injection, robustness of software services, web services interoperability, services security, and enterprise application integration. He has contributed, as an author, reviewer and program committee member, to leading conferences and journals in the dependability and services computing areas. Nuno has been involved in the organisation of several international events, including multiple editions of the International Symposium on Software Reliability Engineering, the Dependable and Secure Services Workshop/Track (as main chair) jointly organised with the IEEE World Congress on Services. He participated in international research projects, including several H2020 projects (e.g., ADVANCE, DEVASSES, ATMOSPHERE, EUBrasilCloudFORUM) and FP7 projects (CRITICAL STEP, CECRIS), and he is currently mostly involved in developing new techniques towards more reliable cloud systems, developing techniques for evaluating the reliability and security of blockchain smart contracts, and using machine learning techniques for software fault and vulnerability detection.
\end{IEEEbiography}

\end{document}